\documentclass{IEEEtran}
\usepackage[T1]{fontenc}
\usepackage[latin9]{inputenc}
\usepackage{color}
\usepackage{array}
\usepackage{float}
\usepackage{booktabs}
\usepackage{tipa}
\usepackage{amsmath}
\usepackage{amssymb}
\usepackage{graphicx}

\makeatletter

\providecommand{\tabularnewline}{\\}

\usepackage{amsmath,amsfonts}

\usepackage{array}

\usepackage{textcomp}
\usepackage{stfloats}
\usepackage{url}
\usepackage{verbatim}
\usepackage{graphicx}

\def\BibTeX{{\rm B\kern-.05em{\sc i\kern-.025em b}\kern-.08em
   T\kern-.1667em\lower.7ex\hbox{E}\kern-.125emX}}
\usepackage{balance}

\usepackage{mathptmx}

\usepackage[plainruled]{algorithm2e}
\usepackage{amsmath}
\usepackage{float}
\floatstyle{plaintop}
\restylefloat{table}
\usepackage{algpseudocode,caption}

\usepackage{graphicx}

\usepackage{subcaption}

\usepackage{txfonts}

\usepackage{lipsum}

\usepackage[numbers,square, comma, sort&compress]{natbib}

\usepackage{babel}

\usepackage{ltxnew}
\usepackage{newtxtt}

\usepackage{pdfcolmk}
\usepackage{varioref}

\usepackage[math]{blindtext}
\usepackage{listings}

\usepackage{mathtools, cuted}

\usepackage{caption}
\makeatother

\begin{document}
\title{Balanced Performance Between Energy-Delay and Bit Error Rate in UAV
Relay Networks }
\author{M khalil, \textit{Member, IEEE}\linebreak{}
}
\maketitle
\begin{abstract}
This paper presents a new strategy for simultaneously reducing energy
consumption, transmission delays, and bit error rate in Unmanned Aerial
Vehicle (UAV) networks. A UAV is fitted with a wireless Bidirectional
Relay (BR) to enable coverage network extension and increase transmission
throughput. The downside of the BR advantages is the delay in data
transmission caused by the UAV\textquoteright s movement. A consequence
of this delay is increased total energy consumption, causing further
degradation in bit error rate performance, especially at high SNR
levels. In wireless communication, the trade-off between delay and
energy consumption is, fortunately, possible to improve performance.
Therefore, this study aims to enhance UAV network performance by reducing
energy consumption, data transmission delay, and bit error rate. A
multi-objective algorithm is employed to generate an adaptive optimal
energy allocation strategy based on the balance between energy consumption
and transmission delays. The results of theoretical analysis are illustrated
with several examples. As herein demonstrated, the proposed solution
effectively balances delay and energy efficiency in a customised system
design and improves the bit error rate in UAV networks.
\end{abstract}


\section{Introduction}

Unmanned aerial vehicles relay network is a communication system employing
drones fitted with wireless relay devices to enhance the scope and
flexibility of communication systems. UAV is currently one of the
essential technologies serving various applications besides telecommunications,
such as product deliveries, aerial photography, policing and surveillance,
military and infrastructure inspections. An example of UAV's importance
has been illustrated in the recent outbreak of Coronavirus disease,
where UAV was employed to perform surveillance public and persuade
them to follow public health best practices \cite{Khalil2020}. In
a telecommunication application, UAV enables communications between
many land user nodes to take place through UAV nodes. It is recognized
that by fitting Decode-and-Forward (DF) or Amplify-and-Forward (AF)
relay systems, the drone network gives better coverage and better
throughput and energy performance \cite{Ono2016}. An AF relay is
considered in this paper as it can theoretically be applied with less
complexity than DF relay, which requires complete processing, including
encoding, re-modulating and re-transmitting the received signal.\textbf{
}Such operations processes require sophisticated power control, which
is unnecessary in an AF relay.\textbf{ }The relay node operates as
either one-way (unidirectional), or two-way (bidirectional). This
paper focuses on the BR AF type as several studies, such as \cite{Zeng2019}
have observed that the BR AF system analysis is appropriate for nano-satellite
communication applications, which are expected to be an essential
part of the 5G networks.

The drone\textquoteright s battery unit, named Battery Eliminator
Circuit (BEC), supplies the necessary energy for the drone, the AF
relay and other components equipped on the drone. Using AF relay nodes
increases the drain of AR\textquoteright s BEC, particularly in the
transmitting mode at which the relay node becomes active. In the transmission
mode, the energy model presents energy consumption for each transmitted
bit. Another factor that increases the drone networks' energy consumption
is the delay due to the UAV movement. Further, transmission signals
in tow-hop significantly impact each node's energy consumption \cite{Zhang2008}.
Rising energy consumption again degrades bit error rate, particularly
at high levels of signal-to-noise ratio (\textit{SNR}) \cite{Zhou2011}.\textbf{
}Moreover, in UAV networks, users and drone nodes often rely on a
battery with a limited amount of energy. Thus, minimizing energy consumption
and delay is a fundamental goal in UAV networks. 

Unfortunately, the delay, energy and bit error rate objectives cannot
be minimized simultaneously as these metrics having conflict with
each other. Such a conflicting problem, however, can be addressed
by using a multi-objective solution \cite{Zappone2018}. One of the
effective methods in multi-objective is the weight of the scalarization
method \cite{Brandt1998}. All objective functions in the weight scalarization
method are consolidated into a single part showing a linear function.
Thus, we can combine the delay, energy and bit error rate objectives
into one function. However, solving such a single function for a bidirectional
relay network is challenging as the communication scenario coincides
in two directions. This paper proposes dividing the problem into two
sub-problems to simplify the weight scalarization function analysis
in a bidirectional relay network. In the first problem, energy consumption
and transmission delay are combined in one function. The solution
to the first problem is employed to optimize energy allocation parameters.
Such parameters reduce delay and energy consumption and enhance bit
error rate. 

Many studies have analysed the balance between energy and delay by
using information theory; it significantly focuses on designing power
allocation under various constraints on the information delay, such
as an average delay constraint for a buffer, queuing delay in \cite{Nuggehalli2002},
per-packet delay constraint \cite{XiliangZhong2005}, in addition,
a multipacket transmission \cite{UysalBiyikoglu2002}. Study \cite{Zafer2005}
uses a minimum departure time as a model of packet delay constraints;
this scheme can be applied to model various quality of service (QoS)
constraints. Further, the design of the system algorithm depends on
the availability of Channel State Information (CSI), which can include
fading channels and time-variation as

According to Shannon's capacity theorem the minimum of $\mathrm{\frac{energy\:per\:bit\:rate(\mathcal{E\mathrm{)}}}{noise\:power\:density\:(\mathcal{N}\mathit{o})}}$
needed to achieve arbitrarily low bit error probability as $\mathcal{\Bigl(E\text{/}\mathrm{\mathcal{N}\mathrm{\mathit{o}}}\Bigr)_{\mathit{\mathrm{min}}}\mathrm{=}}\ln(2).$
For a given error probability and code rate with finite bits, the
minimum energy has been studied in \cite{Shannon1959}. Authors \cite{Polyanskiy2010}
sought to maximize the average throughput, which is the equivalent
of minimizing the average delay-per-bit for a given number of bits
and input power. Study \cite{Polyanskiy2011} presents a minimal energy
solution for transmuting finite bits without delay constraints, and
\cite{Berry2002} a solution for Energy-delay balance over fading
channels has been demonstrated in \cite{Berry2002}. In \cite{Waqar2014},
the model to minimize the energy consumption of the intermediate relay
between the source and the receiver was\textbf{ }adopted for wireless
terrestrial relay. The locations of wireless relays may, however,
be random in practical networks. Hence, in \cite{Waqar2014} the effects
of randomly positioned relays were examined, but this study investigated
selecting the best relay location based on linear places between the
source and destination.\textbf{ }However, in these analyses, the CSI
is assumed to be available at the source or destination nodes.\textbf{
}Nonetheless, the fading channel at mobility relay locations often
varies rapidly, which can make it difficult to estimate, especially
if the relay is moving in space.\textbf{ }

For this reason, Doppler effects are considered by \cite{Zeng2016},
who assumed the relay was flying at a constant speed and the destination
node able to estimate and compensate. The mentioned studies, however,
were focused solely on transmission delay, without taking into consideration
network energy consumption. In contrast, other researchers \cite{Ono2016,Chen2018}
were mainly focused on network energy consumption. Studies \cite{Jiang2018,Zappone2018}
investigated the energy consumption with delay constraints for UAV
AF relay network. Based on the results of these studies, power allocation
is essential\textbf{ }to maximize UAV throughput. However, in many
applications, particularly for energy-limited appliances like sensor
or relay systems, energy is the main parameter that carries out a
specific operation than its power consumption. Generally, the energy
parameter is subject to time delay, and in the meantime, the UAV network
has strict delay regarding safety information transmission delays.
Despite that, \cite{Zappone2018} recognized that distributing energy
value among terrestrial networks nodes reduces delay and overall energy
consumption. 

The location of UAV changes periodically, then the received \textit{SNR}
has time-varying characteristics; thus, bit rate often changes \cite{Govindan2011}.
Accordingly, several researchers worked on determining the effect
of channel characteristics on bit error rate performance. For instance,
study \cite{Ono2016} adopted variable rate protocol to enhance bit
error rate performance and achievable information rate, as the location
of relay changes periodically. Another researcher \cite{Chen2018}
investigated how to place UAVs to reduce bit error rates optimally.
One paper \cite{AlHourani2014} proposed a path loss model that accommodates
both Line of Sight (LOS) and Non-Line of Sight (NLOS) path loss conditions.
Likewise, the authors of \cite{Alzenad2017} extended their results
to include three-dimensional space. In \cite{2016Mozaffar2}, the
optimum location of device-to-device communications was also considered
in UAV to enhance network performance. However, optimising the bit
error rate by selecting the best relay location is impractical. Therefore,
another work \cite{Zhang2018} optimised trajectory and energy control
at the same time. The UAV trajectory is also optimised jointly with
the device-UAV association and uplink power to minimise the total
transmission power according to the number of updates in \cite{Mozaffari2017a}.

All of the aforementioned studies demonstrated effective schemes to
improve UAV networks\textquoteright{} performance in terms of UAV
placement and energy allocation. However, an essential factor that
is largely ignored in these works is that UAV networks may have slightly
larger transmission times, so the data received from the ground user
will have various \textit{SNR} levels. Furthermore, works that developed
the energy consumption and data transmission delay metrics, jointly
or individually, ignored the relationship between these metrics and
the bit error rate, which is an essential metric for evaluating the
performance of UAV network applications \cite{Luo2012}. Thus, \cite{Khalil2021}
presented a specific system model that can only provide balance energy
and delay transmission data for a unidirectional UAV AF relay flying
in a triangle formation. Error-free reception, however, is very limed
in the practical environment, particularly in multi-hop networks with
varying channel conditions. Also, the unidirectional relay operates
in one-way communication between the source and destination, whereas
such communication has limited applicability. To respond to this gap,
this paper considers a general trade-off energy-delay scenario that
enables energy allocation to be an adaptive factor for providing an
optimal bit error rate for UAV bidirectional AF relay networks over
mobile fading channels. The proposed method is achieved by simultaneously
optimising both energy consumption and transmission delay in UAV networks.
The optimum energy allocation is distributed between relay and source
nodes in UAV networks. Such energy allocation solves two problems.
First, it enables the development of a decision-making platform to
achieve the best trade-off between energy consumption and data transmission
delay. Second, it enhances the performance of UAV networks in terms
of bit error rate. Hence, the proposed method simultaneously improves
energy consumption, data transmission delays and bit error rate in
UAV networks. The contribution of this paper is summarised by defining
an algorithm allowing to compute an optimal energy allocation to each
node (users and relay) of the UAV communication network, and further,
it enhances the bit error rate. Most notably, the algorithm strikes
a balance between transmission delays and energy as follows: 
\begin{itemize}
\item Define UAV network by three nodes, which are two terrestrial users
$S_{a}$ and $S_{b}$ and a wireless relay ($\mathit{R}$) fitted
on a drone.
\item Calculate a flight distance ($d_{a}$) between $S_{a}$ and $\mathit{R}$
and the path ($d_{b}$) between $S_{b}$ and $\mathit{R.}$
\item Consider the distances $d_{a}$ and $d_{b}$ to calculate the end-to-end
$\mathit{SNRs}$ at $S_{a}$ and $S_{b}$.
\item Use expressions of $\mathit{SNRs}$ at $S_{a}$ and $S_{b}$ to calculate
energy allocations for $S_{a}$ as $(\alpha_{a}),$ relay $(\alpha_{r})$
and $S_{b}$ as $(\alpha_{b}).$ 
\item Define the bit energy consumption $(\mathcal{E}_{\mathit{a}})$ for
$S_{a}$ and $(\mathcal{E}_{\mathit{b}})$ for $S_{b}$ as functions
of energy allocations factors, i.e., $\mathcal{E}_{\mathit{a}}\mathcal{\mathrm{(\alpha_{a},\alpha_{b},\alpha_{r})}}$
and $\mathcal{E}_{\mathit{b}}\mathcal{\mathrm{(\alpha_{a},\alpha_{b},\alpha_{r})}},$
also
\item Define bit transmission times $(q_{a})$ at $S_{a}$ and $(q_{b})$
at $S_{a}$ as functions of energy allocations as $q_{a}\mathrm{(\alpha_{a},\alpha_{b},\alpha_{r})}$
and $q_{b}\mathrm{(\alpha_{a},\alpha_{b},\alpha_{r})}.$
\item Use Scalarization optimization method demonstrated in \cite{Weck2004}
to optimize $\mathcal{E}_{\mathit{a}}\mathcal{\mathrm{(\alpha_{a},\alpha_{b},\alpha_{r})}}$,
$\mathcal{E}_{\mathit{b}}\mathcal{\mathrm{(\alpha_{a},\alpha_{b},\alpha_{r})}},$
$q_{a}\mathcal{\mathrm{(\alpha_{a},\alpha_{b},\alpha_{r})}}$ and
$q_{b}\mathcal{\mathrm{(\alpha_{a},\alpha_{b},\alpha_{r})}},$ simultaneously,
and this provides an effective method to balance delay-energy performance
on customized system design. In other words, the proposed method offers
a decision-making scheme responsible achieving the best trade-off
between transmission delay and energy consumption in UAV networks. 
\item Use optimum $\alpha_{a},\alpha_{b},\alpha_{r}$ to maximise end-to-end
$\mathit{\textit{\textit{SNR}}}$ and thus lower bit error rates. 
\end{itemize}
The proposed method has been evaluated for the UAV network, and the
analytic outcomes reveal that the proposed approach enables energy
consumption, transmission delay and bit error rate to be minimized
in a well-balanced scheme.

Though the energy allocation strategy has been used to optimise wireless
network performance, as in \cite{Tian2019}, this paper emphasises
the potential of using such a strategy to simultaneously improve the
data rate, energy consumption and bit error rate performance in UAV
networks. According to the author\textquoteright s knowledge, a UAV
network has not been optimised using such a schema previously.

Before further discussion, we provide Table \ref{tab: Table-1} to
summarized notations used in this manuscript.

The rest of this paper is arranged as follows: Sections \ref{sec:SYSTEM-MODEL}
and \ref{Sec 2.2} describe the system model and the received signal
at destination node analysis, respectively; Section \ref{sec:Energy-allocation}
introduces a new method to calculate energy allocation for UAV users
and relay; Section \ref{sec:Simulation-Results} presents the simulation
results; and finally, the conclusion is presented in Section \ref{sec:Conclusion}.

\section{SYSTEM MODEL\label{sec:SYSTEM-MODEL}}

\begin{figure}
\centering{}\includegraphics[width=3.5in]{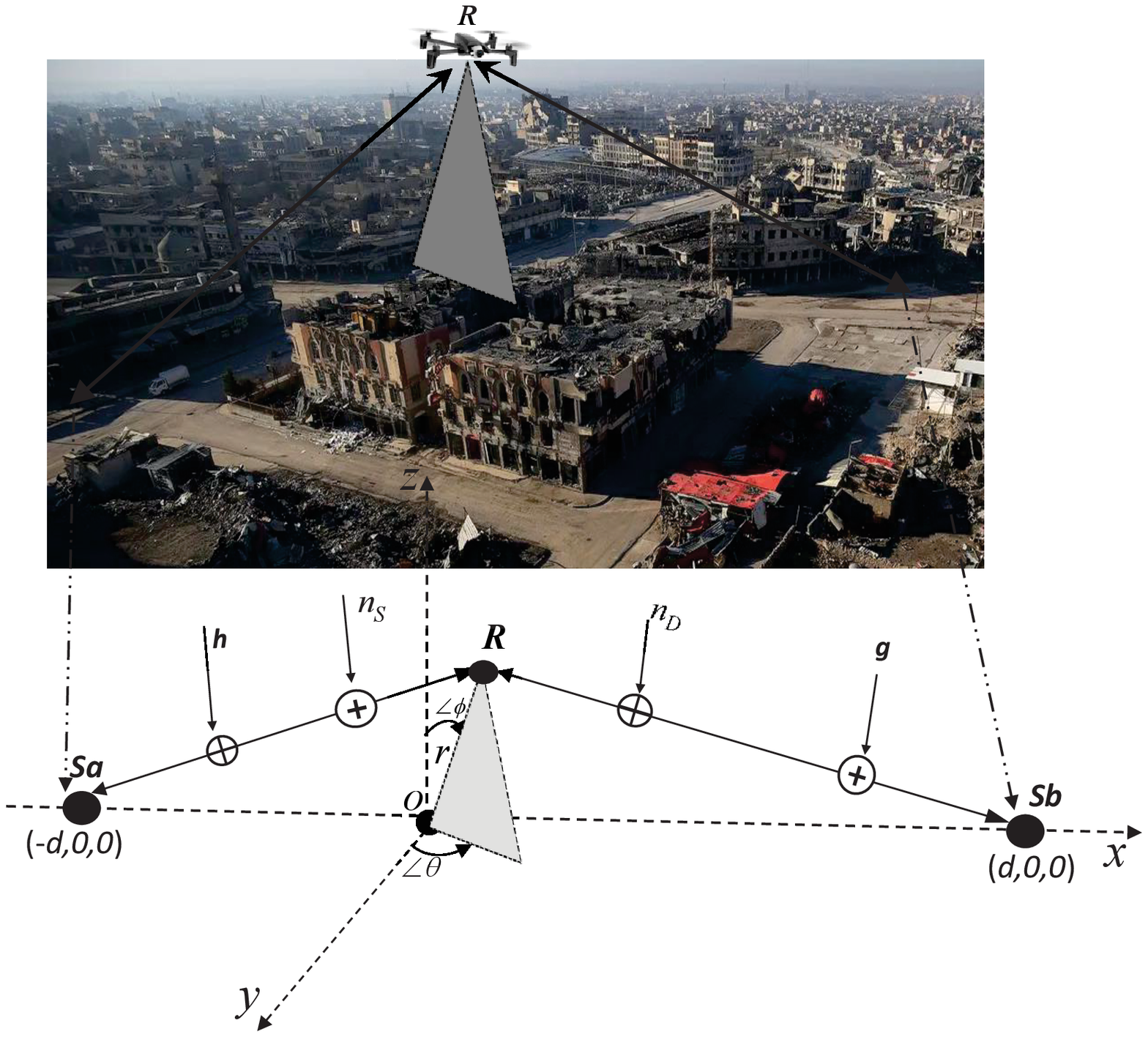}\caption{System Model\label{fig: Fig1}}
\end{figure}

Consider two terrestrial wireless users, $S_{a}$ and $S_{b}$, exchanging
information simultaneously via a wireless relay ($\mathit{R}$) node
fitted with an  drone, i.e., an UAV network. The relay acts in a two-way
amplify-and-forward (AF) relay mode. It is assumed that all the nodes
are equipped with a single antenna, and all the links among the nodes
are half-duplex and use the same carrier frequency. Flying a drone
involves flight along any path in coordinates of three dimensions
($x$, $y$, $z$), where ($\pm x,\,0,\,0$) represent the locations
of land nodes i.e. \textbf{$S_{a}$ }and \textbf{$S_{b}$} and $z$
represents the altitude ($\mathcal{H})$ as shown in Fig \eqref{fig: Fig1}.

For analyzing UAV flight path, we suppose that the drone flies along
a path defined by a series of waypoints, which is assumed to be the
drone\textquoteright s initial position, defined as $\mathrm{p_{i}}$$(\mathrm{\mathit{x}_{i},\,\mathit{y}_{i}\,,\mathit{z}_{i}}).$
Thus, the Cartesian coordinates of $\mathrm{p_{i}}$ are determined
from the coordinate transformations as $\mathrm{\mathit{x}_{i}}=r\cos(\theta)$,
$\mathrm{\mathit{y}_{i}}=r\,\sin(\theta$) and $\mathrm{\mathit{z}_{i}}=r\,\sin(\phi)$.
The polar coordinates $r$, $\theta$ and $\mathcal{\phi}$ are measured
from the centre $O$$(0,0,z)$ of the flight path. 

The Euclidean distance among $S_{a}$, $O$ and $\mathit{S_{b}}$
are calculated at $r=0$, as $\mathit{\pm d}$ which represents the
separating distance between $S_{a}$ and $S_{b}$ . On the other hand,
when $r>0$, the total distance ($d)$ of $d_{a}$ and $d_{b}$ is
varied based on $r$ value. Thus, the Euclidean distance between $S_{a}$
and $\mathit{R}$ is calculated as \vspace{-1em} 

\begin{equation}
d_{a}=\left(d^{2}+\frac{r^{2}}{2}\left(3-\cos(2\phi)\right)-\psi\right)^{1/2},\label{eq: d1}
\end{equation}

and the distance between\textbf{ $R$} and $S_{b}$ is \vspace{-1em} 

\begin{equation}
d_{b}=\left(d^{2}+\frac{r^{2}}{2}\left(3-\cos(2\phi)\right)+\psi\right)^{1/2},\label{eq: d2}
\end{equation}

where $\psi=2\,r\,d\,cos\theta$. 

\begin{table}[H]
\caption{Summary of Notations\label{tab: Table-1}}

\begin{tabular}{p{4cm}p{4cm}}
\toprule 
\addlinespace[0.2cm]
Notations & Description\tabularnewline\addlinespace[0.2cm]
\midrule
\addlinespace[0.2bp]
\midrule 
\addlinespace[0.2cm]
$S_{a}$ and $S_{b}$ & two terrestrial wireless users\tabularnewline\addlinespace[0.2cm]
\addlinespace[0.2bp]
\midrule 
\addlinespace[0.2cm]
$d_{a}$ and $d_{b}$ & the distances between source-drone and drone-receiver, respectively.\tabularnewline\addlinespace[0.2cm]
\addlinespace[0.2bp]
\midrule 
\addlinespace[0.2cm]
$p_{a}$,$p_{b}$ and $p_{R}$ & allocation powers for $S_{a},S_{b}$ and relay, respectively.\tabularnewline\addlinespace[0.2cm]
\addlinespace[0.2bp]
\midrule 
\addlinespace[0.2cm]
$\Re_{a},\Re_{b}$ & data rates for $S_{a},S_{b}$\tabularnewline\addlinespace[0.2cm]
\addlinespace[0.2bp]
\midrule 
\addlinespace[0.2cm]
$h$ and $g$ & fading channels for $d_{a}$ and $d_{b}$ , respectively.\tabularnewline\addlinespace[0.2cm]
\addlinespace[0.2bp]
\midrule 
\addlinespace[0.2cm]
$Y_{a}$ and $Y_{b}$ & received signal by $S_{b}$ and $S_{b}$ , respectively.\tabularnewline\addlinespace[0.2cm]
\addlinespace[0.2bp]
\midrule 
\addlinespace[0.2cm]
$\gamma_{a}$ and $\gamma_{b}$ & SNR at $S_{b}$ and $S_{b}$.\tabularnewline\addlinespace[0.2cm]
\addlinespace[0.2bp]
\midrule 
\addlinespace[0.2cm]
$\mathit{q_{a}}$ and $\mathit{q_{b}}$ & data transmission time by $S_{b}$ and $S_{b}$, respectively.\tabularnewline\addlinespace[0.2cm]
\addlinespace[0.2bp]
\midrule 
\addlinespace[0.2cm]
$\mathcal{E}_{\mathit{a}}$ and $\mathcal{E}_{\mathit{b}}$ & data transmission energies for $S_{b}$ and $S_{b}$, respectively.\tabularnewline\addlinespace[0.2cm]
\addlinespace[0.2bp]
\midrule 
\addlinespace[0.2cm]
$\alpha_{a},$$\alpha_{b}$ and $\alpha_{r}$ & energies allocation factors for $S_{b}$, $S_{b}$ and the relay,
respectively.\tabularnewline\addlinespace[0.2cm]
\addlinespace[0.2bp]
\midrule 
$w_{a}$, $w_{b}$ and $w_{r}$ & weight coefficients for $S_{b}$, $S_{b}$ and the relay, respectively.\tabularnewline
\midrule 
$\text{\texthtb}_{e}$ & bit error rate\tabularnewline
\bottomrule
\end{tabular}
\end{table}

\section{Received signal at detestation node\label{Sec 2.2} }

This part derives an expression for the users $\mathit{SNRs}$ of
the proposed UAV network. We first assume that the destination\textbf{
}is entirely compensated by the Doppler effect due to the UAV\textquoteright s
mobility as the UAV follows a trajectory with a fixed flying speed
\cite{Yang2018}. Thereby, both flat fading channels $\mathit{S_{a}-R}$
and $\mathit{R-S_{b}}$ have power gains following the free-space
path attenuation schemas $h\,d_{a}^{\frac{-\text{\textramshorns}}{2}}$and
$g\,\,d_{b}^{\frac{-\text{\textramshorns}}{2}},$respectively, where
$h$ is the channel coefficient between $S_{a}$ and $R$, $g$ is
the channel coefficient between $S_{b}$ and $R$, $\text{\textramshorns}$
is the path loss exponent which is commonly estimated in the range
of $2\leq\text{\textramshorns}\leq4$. It is also assumed that the
channel\textquoteright s characteristics are available at the user
nodes. Further, user $S_{a}$ is able to adjust the transmit powers
in $S_{b}$ and relay $\mathit{R}$. Similarly, $S_{a}$ can regulate
$S_{a}$ and relay $\mathit{R}$ powers using the power allocation
process.

A further assumption is that the users adopt binary phase shift keying
modulation to broadcast their signals $\chi_{a}$ and $\chi_{b}$,
with average power $p_{a}$ and $p_{b}$, from users $S_{a}$ and
$S_{b}$, respectively. Both signals $\chi_{a}$ and $\chi_{b}$ are
transmitted during the time interval $\mathit{0}<\mathit{t}\leq\mathcal{\mathcal{T}},$and
each signal follows a circularly symmetric complex Gaussian distribution
$\mathcal{CN}(0,1)$, i.e. $E\{\bigl|\chi_{a}\,\bigr|^{2}\}=1$ and
$E\{\bigl|\chi_{b}\bigr|^{2}\}=1$, where $E\{.\}$ denotes an expected
value and $\bigl|.\bigr|^{2}$ is the absolute square of a signal. 

The $S_{a}$ and $S_{b}$ nodes transmit their bits according to a
schedule that defines the commencing and the duration of each bit
transmission. Then, \textbf{$\chi_{a}$ }and\textbf{ $\chi_{b}$ }arrive
to their target node in the time interval $\mathit{0}<\mathit{t}\leq\mathcal{\mathcal{T}}.$
In order to simplify, $\mathcal{T}$ is discretized into $n$ time
slots as $\delta_{\mathit{t}}\mathcal{\mathrm{=}T}$/$\mathcal{N}$,
where $\delta_{\mathit{t}}$ represents the time slot length. The
value of $\delta_{\mathit{t}}$ is assumed to be small enough that
the UAV\textquoteright s location can be supposed to be approximately
constant within a slot. Thereby, the UAV\textquoteright s trajectory\textbf{
$\left(x_{i}\mathrm{(\mathit{t})},y_{i}\mathrm{(\mathit{t})},z_{i}\mathrm{(\mathit{t})}\right)$
}over\textbf{ $\mathcal{T}$ }can be specified\textbf{ }by $\mathcal{N}$
scope as\textbf{ $\bigl\{ x_{i}[n],y_{i}[n],z_{i}\mathrm{(\mathit{n})}\bigr\}_{n}^{\mathcal{N}}$
$\in\mathbb{R^{\mathrm{3}}},$}where\textbf{ }$\mathit{n=1,2,...\mathcal{N-\mathrm{1.}}}$

The exchange signals between $S_{a}$ and $S_{b}$ occur in two-hops.
In the first hop, the relay receives signals from both $S_{a}$ and
$S_{b}$ as\vspace{-1em} 

\begin{equation}
Y_{R}[n]=\sqrt{p_{a}\bigl[n\bigr]}\chi_{a}h+\sqrt{p_{b}\bigl[n\bigr]}\chi_{b\,}g\,d_{b}^{\frac{-\alpha}{2}}+\mathit{n}_{R},\label{eq: yr}
\end{equation}
where $h\,d_{a}^{\frac{-\alpha}{2}}$ and $g\,d_{b}^{\frac{-\alpha}{2}}$
are the fading channels model defined in \cite{Li2011}, $n_{R}$
is the Gaussian noise of the relay with zero mean and variance $(\sigma^{2})$. 

The $Y_{R}$ signal is then amplified by the relay amplification factor
($\beta$) given by:\vspace{-0.75em} 

\begin{equation}
\beta\bigl[n\bigr]=\sqrt{p_{R}\bigl[n\bigr]/(\left|h\right|^{2}d_{a}^{-\alpha}p_{a}\bigl[n\bigr]\,+\left|g\right|^{2}d_{b}^{-\alpha}p_{b}\bigl[n\bigr]+\sigma^{2}}),
\end{equation}
where $p_{R}\bigl[n\bigr]$ is the allocated power for $R$ node . 

After that, each user receives the amplified signals through the same
flat fading channel, during the second time slot. The received signal
by $S_{a}$ is then included in the relay-amplified signal beside
the direct signal received from $S_{b}$ node. This gives \vspace{-2em} 

\begin{gather}
Y_{a}\bigl[n\bigr]=h\,d_{a}^{\frac{-\alpha}{2}}\beta\bigl[n\bigr]y_{R}+\sqrt{p_{b}}\chi_{b\,}g\,d_{b}^{\frac{-\alpha}{2}}\nonumber \\
+2\sqrt{p_{b}}\chi_{b\,}gh\,\left|d\right|^{\frac{-\alpha}{2}}+n_{a}.\label{eq: ya}
\end{gather}

Node $S_{b}$ also involves the relay-amplified signal and the direct
$S_{a}$ signal as the following \vspace{-1em} 

\begin{gather}
Y_{b}\bigl[n\bigr]=g\,d_{b}^{\frac{-\alpha}{2}}\beta\bigl[n\bigr]y_{R}+\sqrt{p_{a}}\chi_{a}\,h\,\,d_{a}^{\frac{-\alpha}{2}}\nonumber \\
+2\sqrt{p_{a}}\chi_{a\,}gh\,\left|d\right|^{\frac{-\alpha}{2}}+n_{b},\label{eq: yb}
\end{gather}

where $gh$ is the channel coefficient between user $S_{a}$ and user
$S_{b}$ with the separating distance 2$\left|d\right|,$and $n_{a}$
and $n_{b}$ are the users Gaussian noises with a zero mean and variance
$(\sigma^{2})$ for both user $S_{a}$ and user $S_{b},$ respectively. 

Equations \eqref{eq: ya} and \eqref{eq: yb} each have their own
transmission signal mixed with the received signal and this is returned
to self-interference (SI). Here, we assume that both users have an
SI cancellation circuit that allows for their own transmitted signal
to be\textbf{ }cancelled out \cite{Khalil2017b}. Thus, the \textit{SNRs}
at each user are given by\vspace{-0.7em}

\begin{align}
\gamma_{a}^{\iota} & \bigl[n\bigr]=\frac{p_{R}\,\gamma_{b}\:p_{b}\,\gamma_{a}\,d_{b}^{-\alpha}d_{h}^{-\alpha}}{\gamma_{a}\:(p_{R}+p_{a})d_{a}^{-\alpha}+p_{b}\,\gamma_{b}d_{b}^{-\alpha}+1}+p_{R}p_{b}\gamma_{ab}\,d_{b}^{-\alpha},\label{eq: SNRa-LOS}\\
\gamma_{b}^{\iota} & \bigl[n\bigr]=\frac{p_{R}\,\gamma_{b}\:p_{a}\,\gamma_{a}\,d_{a}^{-\alpha}d_{b}^{-\alpha}}{p_{a}\,\gamma_{a}\:d_{a}^{-\alpha}+(p_{R}+p_{b})\,\gamma_{b}\,d_{b}^{-\alpha}+1}+p_{R}\,p_{a}\,\gamma_{ab}\,d_{a}^{-\alpha},\label{eq: SNRb-LOS}
\end{align}

where $\gamma_{a}^{\iota}$ and $\gamma_{b}^{\iota}$ are the actual
line-of-sight $\textit{\ensuremath{\mathit{SNR}}}s$ for $S_{a}$
and $S_{b}$, respectively, $\gamma_{ab}=\frac{\left|gh\right|^{2}}{\sigma^{2}},$
$\gamma_{a}=\frac{\left|h\right|^{2}}{\sigma^{2}}$ and $\gamma_{b}=\frac{\left|g\right|^{2}}{\sigma^{2}}.$

In the proposed UAV, the direct link between $S_{a}$ and user $S_{b}$
is usually not possible due to long-distance flight and degradation
in channel quality. Therefore, the destination nodes receive only
the relayed signals from the $R$ node, i.e., $\gamma_{ab}=0$. Equations
\eqref{eq: SNRa-LOS} and \eqref{eq: SNRb-LOS} then become \vspace{-0.7em}

\begin{align}
\gamma_{a_{e}} & \bigl[n\bigr]=\frac{p_{R}\,\gamma_{b}\:p_{b}\,\gamma_{a}\,d_{a}^{-\alpha}d_{b}^{-\alpha}}{\gamma_{a}\:(p_{R}+p_{a})d_{a}^{-\alpha}+p_{b}\,\gamma_{b}d_{b}^{-\alpha}+1},\label{eq: SNRa-NLOS}\\
\gamma_{b_{e}}\bigl[n\bigr] & =\frac{p_{R}\,\gamma_{b}\:p_{a}\,\gamma_{a}\,d_{a}^{-\alpha}d_{b}^{-\alpha}}{p_{a}\,\gamma_{a}\:d_{a}^{-\alpha}+(p_{R}+p_{b})\,\gamma_{b}\,d_{b}^{-\alpha}+1},\label{eq: SNRb-NLOS}
\end{align}
where $\gamma_{a_{e}}$ and $\gamma_{b_{e}}$ are the exact non-line-of-sight
\textit{SNRs} for $S_{a}$ and $S_{b}$, respectively

The average power $\overline{p}_{r}$, $\overline{p}_{a}$ and $\overline{p}_{b}$
should not exceed the overall network power $(\overline{P})$ $\left\{ \overline{P}:\overline{P}\in\mathbb{R},\overline{P}=\underset{i}{\sum}\overline{p}_{i}\quad i\in\left(a,r,b\right)\right\} $.
In order to manage $\overline{P}$ value, the power allocation strategy
is used to allocate a specific power value to the relay and both user
nodes.\textbf{ }All nodes have inclusive knowledge of information
about channels and maximum transmission power for each related node
\cite{Poon2003}, so both $S_{a}$ and $S_{b}$ nodes achieve the
allocation strategy by employing the reverse channels for feedback
allocated energy factors to the corresponding node. After that, $S_{a}$,
$S_{b}$ and relay nodes adjust their transmit power based on these
feedback factors \cite{Zander1997}. 

It assumes that the energy allocation factors $S_{a}$, $\mathit{R}$
and $S_{b}$, respectively, are set as \vspace{-1.6em}

\begin{gather}
\alpha_{a}\Rightarrow\left\{ \alpha_{a}:\in\mathbb{R\mathit{,}}\rightarrow0\leq\alpha_{a}\leq1\right\} ,\label{eq:Alph-1}\\
\alpha_{r}\Rightarrow\left\{ \alpha_{r}:\in\mathbb{R\mathit{,}}\rightarrow0\leq\alpha_{r}\leq1\right\} ,\label{eq:Alph-2}\\
\alpha_{b}\Rightarrow\left\{ \alpha_{b}:\in\mathbb{R\mathit{,}}\rightarrow0\leq\alpha_{b}\leq1\right\} .\label{eq:Alph-3}
\end{gather}

Once received, the corresponding nodes adjust their powers as $\overline{P}\:\alpha_{b},\overline{P}\:\alpha_{r}$
and $\overline{P}\:\alpha_{a}$. So, substituting these regulated
powers into \eqref{eq: SNRa-NLOS} and \eqref{eq: SNRb-NLOS} and
considering high $\mathit{SNR}$ domain, we get\vspace{-1em}

\begin{gather}
\gamma_{a}[n]=\overline{P}\frac{\alpha_{r}\:\bigl[n\bigr]\,\gamma_{b}\:\alpha_{b}\bigl[n\bigr]\,\gamma_{a}d_{a}^{-\text{\textramshorns}}d_{b}^{-\text{\textramshorns}}}{\gamma_{a}\:(\alpha_{r}\:\bigl[n\bigr]\,+\alpha_{a}\:\bigl[n\bigr]\,)d_{a}^{-\text{\textramshorns}}+\alpha_{b}\bigl[n\bigr]\,\gamma_{b}d_{b}^{-\text{\textramshorns}}},\label{eq:Hi-SNR a}\\
\gamma_{b}[n]=\overline{P}\frac{\alpha_{r}\:\bigl[n\bigr]\,\,\gamma_{b}\:\alpha_{a}\:\bigl[n\bigr]\,\,\gamma_{a}d_{a}^{-\text{\textramshorns}}d_{b}^{-\text{\textramshorns}}}{\alpha_{a}\:\bigl[n\bigr]\,\,\gamma_{a}\:d_{a}^{-\text{\textramshorns}}+(\alpha_{r}\:\bigl[n\bigr]\,+\alpha_{b}\:\bigl[n\bigr]\,)\,\gamma_{b}d_{b}^{-\text{\textramshorns}}},\label{eq:HI-SNR b}
\end{gather}

where $\gamma_{a}$ and $\gamma_{b}$ are the high \textit{SNRs} domain
for \eqref{eq: SNRa-NLOS} and \eqref{eq: SNRb-NLOS}, respectively

Now, another metric is the channel capacity, which is given by\vspace{-2em} 

\textbf{
\begin{equation}
C[n]=\frac{1}{2}\log_{2}(1+\phi[n]).\;\quad bits/s,\label{eq: C}
\end{equation}
}where $\phi[n]=1+\gamma_{a}[n]+\gamma_{b}[n]+\gamma_{a}[n]\gamma_{b}[n].$

To adapt reliably transmitted information rate ($\mathcal{R}$), we
have\vspace{-1em}

\begin{equation}
\mathcal{R}[\mathit{n}]=\mathfrak{h}\:C[n]\;\left(0<\text{\text{\ensuremath{\mathfrak{h}}}}<1\right)\label{eq:R}
\end{equation}

At $\text{\text{\ensuremath{\mathfrak{h}}}}=1,$the maximum data rate
in \eqref{eq: C} becomes\vspace{-1em} 

\begin{equation}
\Re_{i}=\Re_{a}+\Re_{b}=\frac{1}{2}\left(\log_{2}(1+\gamma_{a}[n])+\log_{2}(1+\gamma_{b}[n])\right),\label{eq: R-rate}
\end{equation}
where $\Re_{i}$ is the total data rate in two way \cite{Zhang2010}. 

As mentioned earlier, the data rate $\mathcal{R}[\mathit{n}]$ is
passed through two hops, i.e., two cascade channels. The transmission
rate through two channels is subjected to an Information Cascade (IC)
analysis \cite{Kiely1993}, which is defined as a propagation sequence
of data bits transmitting over chaotic channels.\textbf{ }Further,
the mutual-information flow along the cascade of channels cannot exceed
each channel individually. Along the cascade channels, the flow mutual
information capacity cannot overpass each channel capacity. Regarding
UAV channels, the first channel delivers a sequence of bits that transmitted
by users nodes to the $\mathit{R}$ node during a time slot $\mathit{n=\mathrm{1}.}$
The $\mathit{R}$ node requires one-slot processing to forward the
received data to the $S_{a}$ and $S_{b}$ node during the second
time slots, i.e. $\mathit{n}$= 2, 3, 4,...$\mathcal{N}$ \cite{Salehkalaibar2017}.
Then, the information-causality constrain is obtained as: \vspace{-1.5em}

\begin{equation}
\underset{\begin{gathered}S_{a}\rightarrow R\rightarrow S_{b}\\
S_{b}\rightarrow R\rightarrow S_{a}
\end{gathered}
}{\underbrace{\sum_{\mathrm{i=2}}^{\mathrm{n}}\overline{R}\mathrm{[\mathit{i}]}}}\leq\underset{\begin{gathered}R\rightarrow S_{a}\\
R\rightarrow S_{b}
\end{gathered}
}{\underbrace{\sum_{\mathrm{i=1}}^{\mathrm{n-1}}\overline{R}\mathrm{[\mathit{i}]}}}.\label{eq: R1 <R2}
\end{equation}
$S_{a}\rightarrow R\rightarrow S_{b}$ and $S_{b}\rightarrow R\rightarrow S_{a}$
represent the transmitted a signal from $S_{a}$ to $S_{b}$ and from
$S_{b}$ to $S_{a}$, respectively, $R\rightarrow S_{a}$ and $R\rightarrow S_{b}$
typify the amplified signal forwarded by $\mathit{R}$ to $S_{a}$
and $S_{b}$ through one channel.

\begin{algorithm} 
  	 
		\caption{Procedure of $\mathit{SNR}$ analysis}
		 
		\KwIn{%
$\overline{P},\textit{x(t)}, \alpha_{a}, \alpha_{b}, \text{and } \alpha_{r}$  	
	}%
{\bf Initialization of parameters}\\  		 
		  		
\While{$t \neq 0$}{%
$d_{a}\gets \eqref{eq: d1}$\\
$d_{b} \gets \eqref{eq: d2}$ \\
$\gamma_{a} \gets E{Y_{a}[n]}$ \Comment{expectation value of \eqref{eq: SNRa-NLOS}} \\  			
$\gamma_{a} \gets E{ Y_{b}[n]}$ \Comment{expectation value of \eqref{eq: SNRb-NLOS}}   	
$p_{a}=\overline{P}\:\alpha_{a}$ ; $p_{b}=\overline{P}\:\alpha_{b}$; $p_{r}=\overline{P}\:\alpha_{r}$\\
$\gamma_{a} \gets \eqref{eq:Hi-SNR a}$\\   		  		    		
$\gamma_{b} \gets \eqref{eq:HI-SNR b}$
}%
 
 		  		\KwOut{%
			$\mathit{SNR}$ 
 
		}%

\end{algorithm}

\section{Energy Allocation \label{sec:Energy-allocation}}

This section will concentrate on an energy allocation problem to manage
energy consumption in the users and relay nodes of UAV networks. Energy
consumption is related to transmission delay \cite{Ren2019}, so the
proposed energy allocation allows total energy consumption and data
transmission time (delay) to be minimised under a balanced approach.

In Shannon's theorem, prolonging the transmission time reduces transmitting
power. Thus, the total energy consumption in UAV networks is minimised
by maximising transmission time. However, increasing transmission
delay in transmitting the information directly affects the user's
service. Hence, transmission time $(q)$ and power networks must be
designed with a trade-off scheme. $\mathit{q}$ is the amount of time
required by a user to send out a single packet of bits; it depends
on the network's bandwidth and length of the packet, as $q$ = (Data
size/bandwidth) (sec). By using low-latency algorithms or low-delay
transmission protocols, data transmission amounts can be adjusted,
and delays can be reduced. Such approaches allow the data transmission
amount to be managed according to the change in delay performance
due to a change in transmission amount. Each data bit delivered at
$\mathit{q}$ can consume energy ($\mathcal{E})$ as $\mathcal{E}=q\:P$.\textbf{
}Then, the total energy model in UAV is defined from \eqref{eq:Hi-SNR a},
\eqref{eq:HI-SNR b} and \eqref{eq:R} as follows \vspace{-1em}

\begin{gather}
\mathcal{E}_{\mathit{a}}\bigl[n\bigr]=q[n]\left(\:(\alpha_{r}\:\bigl[n\bigr]\,+\alpha_{a}\:\bigl[n\bigr]\,)H+\alpha_{b}\bigl[n\bigr]\,G\right)\left(2^{\frac{2}{q[n]}}-1\right),\label{eq: Ea}\\
\mathcal{\mathcal{E}_{\mathit{b}}}\bigl[n\bigr]=q[n]\left(\alpha_{a}\:\bigl[n\bigr]\,\,H+(\alpha_{r}\:\bigl[n\bigr]\,+\alpha_{b}\:\bigl[n\bigr]\,)\,G\right)\left(2^{\frac{2}{q[n]}}-1\right),\label{eq: Eb}
\end{gather}
where $G=\gamma_{b}d_{b}^{-\text{\textramshorns}}$, $H=\gamma_{a}d_{a}^{-\text{\textramshorns}}$.

Concurrent with the information exchanging between $S_{a}$ and $S_{b},$
the overall energy consumption is typically generated by including
\eqref{eq: Ea} and \eqref{eq: Eb}, i.e., $\,\mathcal{E}\bigl[n\bigr]=\,\mathcal{E}_{\mathit{a}}\bigl[n\bigr]+\mathcal{\mathcal{E}_{\mathit{b}}}\bigl[n\bigr]$,
as in \cite{Ono2016}. The region $\mathcal{E}\bigl[n\bigr]$ can
also be specified by the union of all the possible sets of $\left(\mathcal{E}_{\mathit{a}},\mathcal{E}_{\mathit{b}}\right),$
as in \cite{Yang2018}. Hence, if the region $\mathcal{E}\bigl[n\bigr]$
is managed by varying $\alpha_{a},\alpha_{b},\alpha_{r},$then the
union of the two sets of energies in \eqref{eq: Ea} and \eqref{eq: Eb}
are subject to definition 1.

\textbf{{\normalcolor Definition} {\normalcolor {\color{red}}1.}}
 The regions of two sets $\,\mathcal{E}_{\mathit{a}}$ and $\,\mathcal{E}_{\mathit{b}}$
is the collection of all objects that are in either set. Then, the
union of $\,\mathcal{E}_{\mathit{a}}$ and $\,\mathcal{E}_{\mathit{b}}$
is defined as: $\,\bigcup\mathcal{E}_{i}\bigl[n\bigr]=\mathcal{\,E}_{\mathit{a}}\cup\mathcal{\,E}_{\mathit{b}}\left\{ \alpha_{i}:\left(\alpha_{i}\in\mathcal{\,E}_{\mathit{a}}\right)\lor\left(\alpha_{i}\in\mathcal{\,E}_{\mathit{a}}\right)\right\} $$\quad i\in\left\{ a,\,b,r\right\} $.

Now, the total energy consumption can be expressed in regard to $\alpha_{i}$
as the following \vspace{-1em} 

\begin{equation}
\,\mathcal{E}_{i}(\alpha_{i})\bigl[n\bigr]=\,\mathcal{E}_{\mathit{a}}\bigl[n\bigr]+\mathcal{\mathcal{E}_{\mathit{b}}}\bigl[n\bigr]\quad\quad i\in\left\{ a,b,r\right\} .\label{eq: ener-1}
\end{equation}

For each transmitted bit, the value of $q$ can be defined as $q\bigl[n\bigr]=1/\mathcal{R}[n]$,
so from \eqref{eq:R} we have\vspace{-1em} 

\begin{equation}
\mathit{q_{a}}(\alpha_{i})[\mathit{n}]=2\:\log_{2}\left(1+\,G\overline{P}\frac{\alpha_{r}\:\alpha_{b}}{\alpha_{r}+\alpha_{a}\:+\alpha_{b}\,\frac{G}{H}}\right)^{-1},\label{eq: qa}
\end{equation}

\begin{equation}
\mathit{q_{b}}(\alpha_{i})[\mathit{n}]=2\:\log_{2}\left(1+\:G\overline{P}\frac{\alpha_{r}\:\alpha_{a}}{\alpha_{a}\:+(\alpha_{r}\,+\alpha_{b}\,)\frac{G}{H}}\right)^{-1}i\in\left\{ a,\,b,r\right\} ,\label{eq: qb}
\end{equation}

where (.)$^{-1}$ indicates reciprocal action. 

In two-way relay networks, the powers design of relay and users nodes
are defined as $\left(p_{a}[n]+p_{r}[n]+p_{b}[n]\right)\leq P\bigl[n]$
\cite{Khalil2017b}. In this case, the proposed allocated power is
expressed considering \eqref{eq: R1 <R2} as \vspace{-1.5em}

\begin{equation}
\underset{\mathrm{\mathit{\begin{gathered}\mathrm{the\;}\mathrm{total\:transmit\;power\;of}\\
\mathrm{two\;hops}
\end{gathered}
}}}{\underbrace{\sum_{\mathrm{\mathit{\mathrm{n=2}}}}^{\mathit{\mathcal{N}}}\left(p_{a}[n]+p_{r}[n]+p_{b}[n]\right)}}\leq P\bigl[n\bigr].\label{eq: P-min}
\end{equation}
The purpose of this paper is to optimize $\alpha_{i}$ i.e., $\alpha_{a},\alpha_{b}$
and $\alpha_{\mathit{r}}$ in order to regulate the energy consumption
of \eqref{eq: Ea} and \eqref{eq: Eb} and the transmission delay
of \eqref{eq: qa} and \eqref{eq: qb} in a balanced way. Hence, the
optimisation problem can be formulated as follows \vspace{-1em}

\begin{gather}
\mathcal{\mathrm{min}\:\:E}_{a}(\alpha_{i})[n],\label{eq: Min-Ea}\\
\mathcal{\mathrm{min}\:\:E}_{b}(\alpha_{i})[n],\label{eq: Min-Eb}\\
\mathit{\mathcal{\mathrm{min}\:\:}}\left(q_{a}(\alpha_{i}[\mathit{n}]\mathit{+q_{b}}(\alpha_{i})[\mathit{n}]\right),\label{eq:min-q}\\
s.t\qquad\forall n\quad\alpha_{a}[\mathit{n}]+\alpha_{b}[\mathit{n}]+\alpha_{r}[\mathit{n}]\leq1.\label{eq: const}
\end{gather}

All objectives \eqref{eq: Min-Ea}-\eqref{eq:min-q} must be minimized
at once under the constraint of \eqref{eq: const}; however, the issue
is that both \eqref{eq: Min-Ea} and \eqref{eq: Min-Eb} are in contras
to \eqref{eq:min-q}. Multi-objective optimisation techniques, particularly
the weight scalarization method \cite{Brandt1998}, are some of the
most reliable methods for resolving such an issue. In the weight scalarization
method, all of the objective functions are consolidated into a single
function that appears as a linear function. Many studies have adopted
the scalarization approach to optimise different mathematical functions
such as quadratic and logarithmic functions. By employing the scalarization
approach to minimise \eqref{eq: Min-Ea}-\eqref{eq:min-q} under constraint
\eqref{eq: const}, the following expression is obtained \vspace{-1em}

\begin{gather}
F(\alpha_{a},\alpha_{\mathit{b}},\alpha_{r},w)[n]=\sum_{\mathrm{\mathit{\mathrm{n=2}}}}^{\mathit{\mathcal{N}}}\left(w_{a}\mathcal{\:E}_{a}(\alpha_{i})[n]+w_{b}\mathcal{\:E}_{a}(\alpha_{i})[n]\right.\nonumber \\
\left.+w_{r}\:\left(\mathit{q_{a}}(\alpha_{i})[\mathit{n}]+\mathit{q_{b}}(\alpha_{i})[\mathit{n}]\right)\right)\quad\quad\forall n,\label{eq: 1G-Form-1}
\end{gather}

where $w_{a}\left\{ w_{a}:\,\in\mathbb{R},\;0<w_{a}\leq1\right\} $,
$w_{b}\left\{ w_{b}:\,\in\mathbb{R},\;0<w_{b}\leq1\right\} $ and
$w_{r}\left\{ w_{r}:\,\in\mathbb{R},\;0<w_{r}\leq1\right\} .$ 

The weight coefficients are limited to the following constraint: $\sum_{\mathrm{\mathit{i}}}^{\mathit{m}}w_{i}\leq1,$where
$\mathit{m}$ is the number of functions and $i\in\left\{ a,b,r\right\} .$
Equation \eqref{eq: 1G-Form-1} reveals that both $\mathit{q_{a}}(\alpha_{i})[\mathit{n}]$
and $\mathit{q_{b}}(\alpha_{i})[\mathit{n}]$ express as linear with
a single weight coefficient; this is because the proposed system model
assumes that $S_{a}$ and $S_{b}$, are exchanging information simultaneously
with each other through $\mathit{R}.$

The minimum solution of the objective function $F(\alpha_{a},\alpha_{\mathit{b}},\alpha_{r},w)$
is obtained by satisfying the conditions $\begin{bmatrix}{\normalcolor \frac{\partial\left(\alpha_{a},\alpha_{\mathit{b}},\alpha_{r},w\right)}{\partial\alpha_{a}}} & {\normalcolor \frac{\partial\left(\alpha_{a},\alpha_{\mathit{b}},\alpha_{r},w\right)}{\partial\alpha_{b}}} & {\normalcolor \frac{\partial\left(\alpha_{a},\alpha_{\mathit{b}},\alpha_{r},w\right)}{\partial\alpha_{r}}}\end{bmatrix}^{\mathrm{T}}=0$
and $\begin{bmatrix}{\normalcolor \frac{\partial\left(\alpha_{a},\alpha_{\mathit{b}},\alpha_{r},w\right)}{\partial\alpha_{a}\alpha_{b}}} & {\normalcolor \frac{\partial\left(\alpha_{a},\alpha_{\mathit{b}},\alpha_{r},w\right)}{\partial\alpha_{b}\alpha_{r}}} & {\normalcolor \frac{\partial\left(\alpha_{a},\alpha_{\mathit{b}},\alpha_{r},w\right)}{\partial\alpha_{r}\alpha_{a}}}\end{bmatrix}^{\mathrm{T}}>0$.
Thus, it is required to calculate the following equations\vspace{-0.8em} 

\begin{gather}
\frac{\partial F(\alpha_{a},\alpha_{\mathit{b}},\alpha_{r},w)}{\partial\alpha_{a}}=q[n]\;H\;\left(2^{\frac{2}{q[n]}}-1\right)\left(w_{a}+w_{b}\right)\nonumber \\
-w_{r}G\overline{P}\left(\frac{\alpha_{b}H^{2}\alpha_{r}}{\left(\alpha_{b}G+H\alpha_{r}+H\alpha_{a}\right)^{2}}+\frac{GH\alpha_{r}(\alpha_{b}+\alpha_{r})}{\left(\alpha_{b}G+G\alpha_{r}+H\alpha_{a}\right)^{2}}\right).\label{eq:da}\\
\frac{\partial F(\alpha_{a},\alpha_{\mathit{b}},\alpha_{r},w)}{\partial\alpha_{b}}=q[n]\;G\;\left(2^{\frac{2}{q[n]}}-1\right)\left(w_{a}+w_{b}\right)\nonumber \\
+w_{r}G\overline{P}\left(\frac{H^{2}\alpha_{r}(\alpha_{a}+\alpha_{r})}{\left(\alpha_{b}G+H\alpha_{r}+H\alpha_{a}\right)^{2}}-\frac{\alpha_{a}GH\alpha_{r}}{\left(\alpha_{b}G+G\alpha_{r}+H\alpha_{a}\right)^{2}}\right),\label{eq: db}\\
\frac{\partial F(\alpha_{a},\alpha_{\mathit{b}},\alpha_{r},w)}{\partial r}=q[n]\;H\;\left(2^{\frac{2}{q[n]}}-1\right)\left(H\,w_{a}+G\,w_{b}\right)\nonumber \\
+w_{r}G\overline{P}\left(\frac{\alpha_{b}H(\alpha_{a}H+\alpha_{b}G)}{\left(\alpha_{b}G+H\alpha_{r}+H\alpha_{a}\right)^{2}}+\frac{\alpha_{a}H(\alpha_{a}H+\alpha_{b}G)}{\left(\alpha_{b}G+G\alpha_{r}+H\alpha_{a}\right)^{2}}\right).\label{eq:dr}
\end{gather}

Also we have\vspace{-1em} 

\begin{gather}
\begin{bmatrix}{\normalcolor \frac{\partial\left(\alpha_{a},\alpha_{\mathit{b}},\alpha_{r},w\right)}{\partial\alpha_{a}\alpha_{b}}}\\
{\normalcolor \frac{\partial\left(\alpha_{a},\alpha_{\mathit{b}},\alpha_{r},w\right)}{\partial\alpha_{b}\alpha_{r}}}\\
{\normalcolor \frac{\partial\left(\alpha_{a},\alpha_{\mathit{b}},\alpha_{r},w\right)}{\partial\alpha_{r}\alpha_{a}}}
\end{bmatrix}=\nonumber \\
\begin{bmatrix}\frac{-H^{2}r(\alpha_{a}H-G\alpha_{b}+H\alpha_{r})G\overline{P}}{(\alpha_{a}H+G\alpha_{b}+H\alpha_{r})^{3}} & \frac{GHr(\alpha_{a}H-G(\alpha_{r}+\alpha_{b}))G\overline{P}}{(\alpha_{a}H+G(\alpha_{r}+\alpha_{b}))^{3}}\\
\frac{G\overline{P}H^{2}\left(\alpha_{a}^{2}H+\alpha_{a}bG+aH\alpha_{r}+2\alpha_{b}G\alpha_{r}\right)}{(H(\alpha_{a}+\alpha_{r})+\alpha_{b}G)^{3}} & \frac{-\alpha_{a}GH(\alpha_{a}H+\alpha_{b}G-G\alpha_{r})G\overline{P}}{(\alpha_{a}H+\alpha_{b}G+G\alpha_{r})^{3}}\\
\frac{\alpha_{b}H^{2}(\alpha_{b}G+H(\alpha_{a}-\alpha_{r}))G\overline{P}}{(\alpha_{b}G+H(\alpha_{r}+\alpha_{a}))^{3}} & \frac{2\alpha_{a}H^{2}(\alpha_{a}H+\alpha_{b}G)G\overline{P}}{(\alpha_{b}G+G\alpha_{r}+H\alpha_{a})^{3}}
\end{bmatrix}.\label{eq: second drive}
\end{gather}

Equation \eqref{eq: second drive} illustrates that the second partial
derivative test $\begin{bmatrix}{\normalcolor \frac{\partial\left(\alpha_{a},\alpha_{\mathit{b}},\alpha_{r},w\right)}{\partial\alpha_{a}\alpha_{b}}} & {\normalcolor \frac{\partial\left(\alpha_{a},\alpha_{\mathit{b}},\alpha_{r},w\right)}{\partial\alpha_{b}\alpha_{r}}} & {\normalcolor \frac{\partial\left(\alpha_{a},\alpha_{\mathit{b}},\alpha_{r},w\right)}{\partial\alpha_{r}\alpha_{a}}}\end{bmatrix}^{\mathrm{T}}>0$.
Thus, the minimum local points of equations \eqref{eq:da} to \eqref{eq:dr}
are obtained after setting the first conditions, i.e., $\frac{\partial F(\alpha_{a},\alpha_{\mathit{b}},\alpha_{r},w)}{\partial\alpha_{a}}=0$,
$\frac{\partial F(\alpha_{a},\alpha_{\mathit{b}},\alpha_{r},w)}{\partial\alpha_{b}}=0$
and $\frac{\partial F(\alpha_{a},\alpha_{\mathit{b}},\alpha_{r},w)}{\partial\alpha_{r}}=0$.

Before continuing with this analysis, another test should be applied
to $F\mathit{\mathrm{[}n}\bigr](\alpha_{a},\alpha_{\mathit{b}},\alpha_{r})$.
This test aims to find whether the $F$ function is a convex function
or not, as the weigh scalarization method is inefficient to find a
solution with a non-convex function \cite{Koski1985}. To achieve
such a test,\textbf{ }the Bordered Hessian ($\mathit{H_{b}}$) matrix
is employed as follows: \vspace{-1em}

\begin{gather}
\mathtt{z}^{\mathtt{T}}\mathit{H_{b}}\mathtt{z}=\begin{bmatrix}\mathtt{z_{1}} & \mathtt{z_{2}} & \mathtt{z_{3}} & \mathtt{z_{4}}\end{bmatrix}\nonumber \\
\left[\begin{array}{ccc}
0 & \frac{\partial F\left(\alpha_{a},\alpha_{\mathit{b}},\alpha_{r},w\right)}{\partial\alpha_{a}} & \frac{\partial F\left(\alpha_{a},\alpha_{\mathit{b}},\alpha_{r},w\right)}{\partial\alpha_{b}}\\
\frac{\partial F(\alpha_{a},\alpha_{\mathit{b}},\alpha_{r},w)}{\partial\alpha_{a}} & \frac{\partial^{2}F(\alpha_{a},\alpha_{\mathit{b}},\alpha_{r},w)}{\partial\alpha_{a}^{2}} & \frac{\partial F^{2}(\alpha_{a},\alpha_{\mathit{b}},\alpha_{r},w)^{2}}{\partial\alpha_{a}\partial\alpha_{b}}\\
\frac{\partial F(\alpha_{a},\alpha_{\mathit{b}},\alpha_{r},w)}{\partial\alpha_{b}} & \frac{\partial^{2}F(\alpha_{a},\alpha_{\mathit{b}},\alpha_{r},w)}{\partial\alpha_{a}\partial\alpha_{b}} & \frac{\partial^{2}F(\alpha_{a},\alpha_{\mathit{b}},\alpha_{r},w)}{\partial\alpha_{b}^{2}}\\
\frac{\partial F(\alpha_{a},\alpha_{\mathit{b}},\alpha_{r},w)}{\partial\alpha_{r}} & \frac{\partial^{2}F(\alpha_{a},\alpha_{\mathit{b}},\alpha_{r},w)}{\partial\alpha_{a}\partial\alpha_{r}} & \frac{\partial^{2}F(\alpha_{a},\alpha_{\mathit{b}},\alpha_{r},w)}{\partial\alpha_{b}\partial\alpha_{r}}
\end{array}\right.\nonumber \\
\left.\begin{array}{c}
\frac{\partial F\left(\alpha_{a},\alpha_{\mathit{b}},\alpha_{r},w\right)}{\partial\alpha_{r}}\\
\frac{\partial^{2}F(\alpha_{a},\alpha_{\mathit{b}},\alpha_{r},w)}{\partial\alpha_{a}\partial\alpha_{r}}\\
\frac{\partial^{2}F(\alpha_{a},\alpha_{\mathit{b}},\alpha_{r},w)}{\partial\alpha_{b}\partial\alpha_{r}}\\
\frac{\partial^{2}F(\alpha_{a},\alpha_{\mathit{b}},\alpha_{r},w)}{\partial\alpha_{r}^{2}}
\end{array}\right]\begin{bmatrix}\mathtt{z_{1}}\\
\mathtt{z_{2}}\\
\mathtt{z_{3}}\\
\mathtt{z_{4}}
\end{bmatrix}.\label{eq: Hessian}
\end{gather}

Substituting \eqref{eq:da}-\eqref{eq: second drive} into \eqref{eq: Hessian}
leads to the following results: $\mathtt{z}^{\mathtt{T}}\mathit{H_{b}}\mathtt{z}\geq0$,
i.e., $\mathit{H_{b}}$ is non-negative. Therefore, the Hessian matrix
is positive semidefinite; hence $F(\alpha_{a},\alpha_{\mathit{b}},\alpha_{r},w)$
is a set of convex. 

Now, we evaluate $\begin{bmatrix}{\normalcolor \frac{\partial\left(\alpha_{a},\alpha_{\mathit{b}},\alpha_{r},w\right)}{\partial\alpha_{a}}} & {\normalcolor \frac{\partial\left(\alpha_{a},\alpha_{\mathit{b}},\alpha_{r},w\right)}{\partial\alpha_{b}}} & {\normalcolor \frac{\partial\left(\alpha_{a},\alpha_{\mathit{b}},\alpha_{r},w\right)}{\partial\alpha_{r}}}\end{bmatrix}^{\mathrm{T}}=0$
from \eqref{eq:da} to \eqref{eq:dr}. This gives \vspace{-1em} 

\begin{gather}
\overset{*}{\alpha}_{a}(w_{j})=\frac{1}{\psi_{a}}\left(\psi_{b}\pm\sqrt{\psi_{b}^{2}+2\psi_{a}\psi_{c}}\right)\quad\forall n,\label{eq:opt-Alph-A}\\
\overset{*}{\alpha}_{\mathit{b}}(w_{j})=\left(\frac{1}{\psi_{a}}\left(\psi_{b}\pm\sqrt{\psi_{b}^{2}+2\psi_{a}\psi_{c}}\right)\left(\frac{H}{G}-1\right)+1\right)\nonumber \\
\sqrt{\frac{q\left(2^{\frac{2}{q[n]}}-1\right)\left[\;H-G\right]}{2HG\overline{P}\left(w_{3}\right)}\left(1-w_{3}\right)}\quad\forall n,\label{eq:opt-Alph-B}\\
\overset{*}{\alpha}_{\mathit{r}}(w_{j})=1-\frac{1}{\psi_{a}}\left(\psi_{b}\pm\sqrt{\psi_{b}^{2}+2\psi_{a}\psi_{c}}\right)\nonumber \\
-\left(\frac{1}{\psi_{a}}\left(\psi_{b}\pm\sqrt{\psi_{b}^{2}+2\psi_{a}\psi_{c}}\right)\left(\frac{H}{G}-1\right)+1\right)\nonumber \\
\sqrt{\frac{q\left(2^{\frac{2}{q[n]}}-1\right)\left[\;H-G\right]}{2HG\overline{P}\left(w_{3}\right)}\left(1-w_{3}\right)}\quad\forall n,\label{eq:opt-Alph-r}
\end{gather}

where $(w_{j}):j\in\left\{ a\vee\,b,r\right\} $, $\vee$ is 'or'
symbol, $\psi_{a}=2\left(G-H\right)$
\begin{align*}
\left(\mathrm{1}-\left(\frac{q\left(2^{\frac{2}{q[n]}}-1\right)\left[H-G\right]\left(w_{2}\right)}{2HG\overline{P}\left(w_{3}\right)}-\frac{1}{\left(G-H\right)+H}\right)\mathrm{(G-H)}\right) & ,
\end{align*}
$\psi_{b}={\color{black}\left(\frac{H}{G}-1\right)}\sqrt{G\frac{q\left(2^{\frac{2}{q[n]}}-1\right)\left[\;H-G\right]}{2H\overline{P}\left(w_{3}\right)}\left(1-w_{3}\right)}$+2G
and $\psi_{c}=\sqrt{G\frac{q\left(2^{\frac{2}{q[n]}}-1\right)\left[\;H-G\right]}{2H\overline{P}\left(w_{3}\right)}\left(1-w_{3}\right)}-G$.

The solution of \eqref{eq:opt-Alph-A}-\eqref{eq:opt-Alph-r} depend
on any variation of two-weight coefficient, and this gives an expected
result because the proposed system model has two sources, $S_{a}$
and $S_{b}$, and both sources contribute to adjusting the energy
allocation parameters. Based on several $w_{j}$, \eqref{eq:opt-Alph-A}-\eqref{eq:opt-Alph-r}
provide a trade-off between energy consumption and transmission time.
This agrees with the analysis in \cite{To1996}, which demonstrated
that the weight scalarization method produce trade-off solution by
repeating the analysis process for several weight coefficients. 

Thus, the solution \eqref{eq:min-q} is obtained by using \eqref{eq:opt-Alph-A}-\eqref{eq:opt-Alph-r}
as follows: \vspace{-2em} 

\begin{equation}
\overset{*}{q_{i}}\bigl[n\bigr](w_{j})=2/\left(1+\log_{2}\left(1+\varPhi\bigl[n\bigr]\right)\right).\label{eq: opt_qt}
\end{equation}

By adding \eqref{eq: Min-Ea} and \eqref{eq: Min-Eb} together, the
total energy consumption solution is calculated in terms of \eqref{eq:opt-Alph-A}-\eqref{eq:opt-Alph-r}
as follows: \vspace{-1em} 

\begin{gather}
\mathcal{\mathcal{\overset{*}{E}}}_{i}\bigl[n\bigr](w_{j})=\overset{*}{q_{i}}\bigl[n\bigr]\left(2^{\frac{2}{q[n]}}-1\right)\left(\left(\overset{*}{\alpha}_{\mathit{r}}(w_{j})\,+2\overset{*}{\alpha}_{a}(w_{j})\right)H\right.\nonumber \\
\left.+\left(\overset{*}{\alpha}_{\mathit{r}}(w_{j})+2\overset{*}{\alpha}_{\mathit{b}}(w_{j})\right)\,G\right)\quad j\in\left\{ a\,\vee\,b,r\right\} ,\label{eq: Opt-Ei}
\end{gather}

where, $\varPhi\bigl[n\bigr]=1+\log_{2}\left(1+\,G\overline{P}\frac{\overset{*}{\alpha}_{\mathit{r}}(w_{j})\:\overset{*}{\alpha}_{\mathit{b}}(w_{j})}{\overset{*}{\alpha}_{\mathit{r}}(w_{j})+\overset{*}{\alpha}_{a}(w_{j})+\overset{*}{\alpha}_{\mathit{b}}(w_{j})\,\frac{G}{H}}\right)$

$+\log_{2}\left(1+\,G\overline{P}\frac{\overset{*}{\alpha}_{\mathit{r}}(w_{j})\overset{*}{\alpha}_{\mathit{b}}(w_{j})}{\overset{*}{\alpha}_{\mathit{r}}(w_{j})+\overset{*}{\alpha}_{a}(w_{j})+\overset{*}{\alpha}_{\mathit{b}}(w_{j})\,\frac{G}{H}}\right)$
$\left(1+\:G\overline{P}\frac{\overset{*}{\alpha}_{\mathit{r}}(w_{j})\:\overset{*}{\alpha}_{a}(w_{j})}{\overset{*}{\alpha}_{a}(w_{j})+(\overset{*}{\alpha}_{\mathit{r}}(w_{j})+\overset{*}{\alpha}_{\mathit{b}}(w_{j}))\frac{G}{H}}\right)$

$+\log_{2}\left(1+\:G\overline{P}\frac{\overset{*}{\alpha}_{\mathit{r}}(w_{j})\:\overset{*}{\alpha}_{a}(w_{j})}{\overset{*}{\alpha}_{a}(w_{j})+(\overset{*}{\alpha}_{\mathit{r}}(w_{j})+\overset{*}{\alpha}_{\mathit{b}}(w_{j}))\frac{G}{H}}\right).$

The \eqref{eq: opt_qt} and \eqref{eq: Opt-Ei} procedures are listed
in Algorithm 2.

\begin{algorithm} 
  		\caption{Proposed energy-delay trade-off}
		 
		\KwIn{%
${\gamma_{a}, \text{and } \gamma_{b}}$   	
	}%
{\bf Initialization of parameters}\\  		 
		  		
\While{$t \neq 0$}{%
$\mathcal{R}\gets \eqref{eq: R-rate}$\\
 $\mathcal{E}_{a}(\alpha_{a},\alpha_{b},\alpha_{r})\gets \eqref{eq: Ea}$\\ 
$\mathcal{E}_{b}(\alpha_{b},\alpha_{b},\alpha_{r})\gets \eqref{eq: Eb}$\\ 
 $\mathit{q}_{a}(\alpha_{a},\alpha_{b},\alpha_{r})\gets \eqref{eq: qa}$\\ 
 $\mathit{q}_{b}(\alpha_{a},\alpha_{b},\alpha_{r})\gets\eqref{eq: qb}$\\ 
\For{$w_{i}=0:$ \KwTo  $1$}{
\For{$w_{j}=0:$ \KwTo  $1$}{

 $F(\alpha_{a},\alpha_{b},\alpha_{r},w_{i,j}) \gets \eqref{eq: 1G-Form-1}$\\  
  $\overset{*}{\alpha_{a}}(w_{i,j}) \gets \eqref{eq:opt-Alph-A}$\\ 
  $\overset{*}{\alpha_{b}}(w_{i,j}) \gets \eqref{eq:opt-Alph-B}$\\   
 $\overset{*}{\alpha_{r}}(w_{i,j}) \gets \eqref{eq:opt-Alph-r}$\\
$\overset{*}{\mathit{q}_{i}}(\overset{*}{\alpha_{a}}, \overset{*}{\alpha_{b}}, \overset{*}{\alpha_{r}}) \gets \eqref{eq: opt_qt}$\\
$\overset{*}{\mathcal{E}_{i}}(\overset{*}{\alpha_{a}}, \overset{*}{\alpha_{b}}  \overset{*}{\alpha_{r}}) \gets\eqref{eq: Opt-Ei}$\\

}	
		}

}%

 		  		\KwOut{%
  $\mathcal{E}_{a}(\alpha_{a},\alpha_{b},\alpha_{r})$;$\mathcal{E}_{b}(\alpha_{b},\alpha_{b},\alpha_{r})$;
$\mathit{q}_{a}(\alpha_{b},\alpha_{b},\alpha_{r})$;$\mathit{q}_{b}(\alpha_{b},\alpha_{b},\alpha_{r})$ 
 
		}%

\end{algorithm}

\section{Bit error rate performance \label{sec:3}}

This section discusses the optimum bit error rate ($\overset{*}{\text{\texthtb}_{e}}$)
behaviour of the UAV network based on Equation \eqref{eq: Opt-Ei}.
The bit error metric defines the number of errors that occur during
data transmission on the UAV network. It can be reformatted for each
transmitted bit (i.e. $\mathcal{\overline{R}}=1)$ as follows 

\begin{equation}
\frac{1}{\overset{*}{\gamma}(w_{j})[n]}=\frac{1}{\left(\left(\overset{*}{\alpha}_{\mathit{r}}(w_{j})\,+2\overset{*}{\alpha}_{a}(w_{j})\right)H\overline{P}\right.}+\,\frac{1}{\left(\overset{*}{\alpha}_{\mathit{r}}(w_{j})+2\overset{*}{\alpha}_{\mathit{b}}(w_{j})\right)\overline{P}G},\label{eq: opt-SNR}
\end{equation}

where $\overset{*}{\gamma}(w_{j})$ the optimal \textit{SNR}.

Equation \eqref{eq: opt-SNR} reveals that the $\mathit{SNR}$ is
increasing function of optimal allocation factor $(\varphi_{i}),$
$\varphi_{i}\in\left(\overset{*}{\alpha_{a}}(w_{j}),\overset{*}{\alpha}_{b}(w_{j}),\overset{*}{\alpha}_{r}(w_{j})\right)$.
Further, according to \cite{Farhadi2008}, bit error rate decreases
when the overall received \textit{SNR} is maximised. Thus, increasing
$\varphi_{i}$ in \eqref{eq: opt-SNR} allows bit error rate to be
minimised as demonstrate in the following analysis.

In the proposed UAV  network, the transmitted signal from a $\mathit{S_{a}}$
to the $\mathit{\mathit{S_{b}}}$ node propagates through two cascaded
channels, as illustrated in \eqref{fig: Fig1}. Each channel is a
Rayleigh fading and, in such a link, \textit{SNR} follows a negative
exponential distribution. Then the total \textit{SNR} obtained from
two i.i.d channels follows a negative exponential distribution. To
find the total probability density function (pdf) of the $\mathit{S_{a}-R}$
and $\mathit{R-S_{b}}$ channels, distribution of the harmonic mean
of two i.i.d. gamma random variables demonstrated in \cite{Hasna2004a}
is applied as follows: first define pdf of the $\mathit{S_{a}-R}$
as\vspace{-1em} 

\begin{equation}
pdf^{h}=\frac{e^{-\frac{\eta}{\gamma_{h^{p}}[n]}}}{\gamma_{h^{p}}[n]},\label{eq: pdf1}
\end{equation}
and pdf of the $\mathit{R-S_{b}}$ link as \vspace{-1em} 

\begin{equation}
pdf^{g}=\frac{e^{-\frac{\eta}{\gamma_{g^{p}}[n]}}}{\gamma_{g^{p}}[n]},\label{eq:pdf2}
\end{equation}
where $\gamma_{h^{p}}[n]=\overset{*}{\alpha}_{\mathit{r}}(w_{j})\:H\overline{P}$,
$\gamma_{g^{p}}[n]=\frac{\overset{*}{\alpha}_{\mathit{r}}(w_{j})\:\overset{*}{\alpha}_{\mathit{b}}(w_{j})}{\overset{*}{\alpha}_{\mathit{r}}(w_{j})+\overset{*}{\alpha}_{a}(w_{j})}G\overline{P}$,
and $\eta$ is the harmonic mean according to a gamma distribution
defined as $\eta=\mu_{H}(\gamma_{h^{p}}[n],\gamma_{g^{p}}[n])$ \cite{Hasna2003a}. 

To joint \eqref{eq: pdf1} and \eqref{eq:pdf2}, the modified harmonic
mean demonstrated in \cite{Louie2009} is applied as follows \vspace{-1em} 

\begin{gather}
pdf^{t}(\eta)=\left(\frac{2}{\gamma_{h^{p}}[n]\,\gamma_{g^{p}}[n]}k_{0}\left(\frac{2\eta}{\sqrt{\gamma_{h^{p}}[n]\,\gamma_{g^{p}}[n]}}\right)\right.\nonumber \\
\frac{\gamma_{h^{p}}[n]+\gamma_{g^{p}}[n]}{\bigl(\gamma_{h^{p}}[n]\gamma_{g^{p}}[n]\bigr)^{3/2}}\:k_{1}\left(\frac{2\eta}{\sqrt{\gamma_{h^{p}}[n]\,\gamma_{g^{p}}[n]}}\right)2\eta\,e^{-\eta\,(\frac{\gamma_{h^{p}}[n]+\gamma_{g^{p}}[n]}{\gamma_{h^{p}}[n]\,\gamma_{g^{p}}[n]})},\label{eq:Bassel}
\end{gather}

where $pdf^{t}(\eta)$ is the total pdf, $k_{0}$$(.)$ and $k_{1}(.)$
are the first and the second order modified Bessel function of the
second kind.

Equation \eqref{eq:Bassel} is simplified by applying the modified
Bessel function properties as $k_{0}(\eta)\rightarrow0$ and $k_{1}(\underset{\eta\rightarrow0}{\eta})\rightarrow1/\eta$.
This gives $pdf^{t}(\eta)\approx\left(\frac{1}{\gamma_{h^{p}}[n]}+\frac{1}{\gamma_{g^{p}}[n]}\right)e^{-\eta\left(\frac{1}{\gamma_{h^{p}}[n]}+\frac{1}{\gamma_{g^{p}}[n]}\right)}.$
By integrating $pdf^{t}(\eta)$ relative to $\eta$, the cumulative
distribution function is obtained \vspace{-1em} 

\begin{equation}
\mathcal{F}(\eta)=1-e^{-\eta\left(\frac{1}{\gamma_{h^{p}}[n]}+\frac{1}{\gamma_{g^{p}}[n]}\right)},\label{eq: CDF1-1}
\end{equation}
where $\mathcal{F}(\eta)$ is the cumulative distribution function
for $pdf^{t}(\eta).$

At high $\mathit{\textit{\textit{SNR}}}$ domain, the first order
expansion of $\mathcal{F}(\eta)$ is given by

\begin{equation}
\mathcal{F}(\eta)=\eta\left(\frac{1}{\gamma_{h^{p}}[n]}+\frac{1}{\gamma_{g^{p}}[n]}\right)+o\left(x^{1+\varepsilon}\right),0<\varepsilon<1\label{eq: exp}
\end{equation}

Now, the approximate bit error rate of the UAV  at a high $\mathit{\textit{\textit{SNR}}}$
can be roughly estimated by using \cite{Wang2003} as \vspace{-1em} 

\begin{equation}
\mathrm{Bit\,error\:rate}=\mathbb{E}\left\{ Q\left(\sqrt{2\gamma}\right)\right\} =\frac{1}{2\sqrt{\pi}}\int_{0}^{\infty}\frac{e^{-\eta}}{\sqrt{\eta}}\mathcal{F}(\eta)\:d\eta.\label{eq:Q(r1)}
\end{equation}

Substituting \eqref{eq: exp} in \eqref{eq:Q(r1)} and evaluating
the result lead to obtaining the optimal bit error rate of the UAV
as

\begin{multline}
\overset{*}{\text{\texthtb}}[n]=\frac{\Gamma\left(\frac{3}{2}\right)}{\overset{*}{\alpha}_{\mathit{r}}(w_{j})\overline{P}\sqrt{\pi}}\\
\left(\frac{1}{\left(\overset{*}{\alpha}_{\mathit{r}}(w_{j})\,+2\overset{*}{\alpha}_{a}(w_{j})\right)H}+\frac{1}{\left(\overset{*}{\alpha}_{\mathit{r}}(w_{j})+2\overset{*}{\alpha}_{\mathit{b}}(w_{j})\right)G}\right),\label{eq: BER}
\end{multline}

where $\Gamma(.)$ is gamma function.

In Algorithm 3, the \eqref{eq: BER} procedure is demonstrated.

\begin{algorithm} 
  		\caption{ Bit error rate performance}
		 
		\KwIn{%
$\overset{*}{\varphi_{i}}$  	
	}%
{\bf Initialization of parameters}\\  		 
		  		
\While{$t \neq 0$}{%
\For{$w_{i}=0:$ \KwTo  $1$}{
$\overset{*}{\gamma}(w_{i})[n]\gets \eqref{eq: opt-SNR}$\\
$pdf^{h} \gets \eqref{eq: pdf1}$\\  

$pdf^{g}\gets \eqref{eq:pdf2}$\\
$\mathcal{F}(\eta)\gets \eqref{eq: exp}$ \\
 }
}%
 
 		  		\KwOut{%
			 $\text{\texthtb}_{e}[n]$

		}%

\end{algorithm}

\section{Simulation Results\label{sec:Simulation-Results}}

In this section, numerical simulations are carried out to assess transmission
delay and energy savings for the proposed UAV network. The analytical
outcomes are validated by utilizing Monte Carlo simulations that use
$10^{5}$ samples.\textbf{ }The rest of the simulation specifications
are set as: $100\leq d\leq700$ meters and the maximum powers for
users and relay nodes are specified by 2 watts. 

The simulation results, which are plotted simultaneously using equations
\eqref{eq: opt_qt}-\eqref{eq: Opt-Ei} refer to the \textit{Proposed
Algorithm $\mathrm{(\mathit{PA})}$} scheme. The results of \eqref{eq: Ea}-\eqref{eq: qb}
are plotted under the name \textit{Sub-optimal Network} $\mathit{\mathrm{(}SN\mathrm{).}}$

\begin{figure}
\centering{}\includegraphics[width=3.5in]{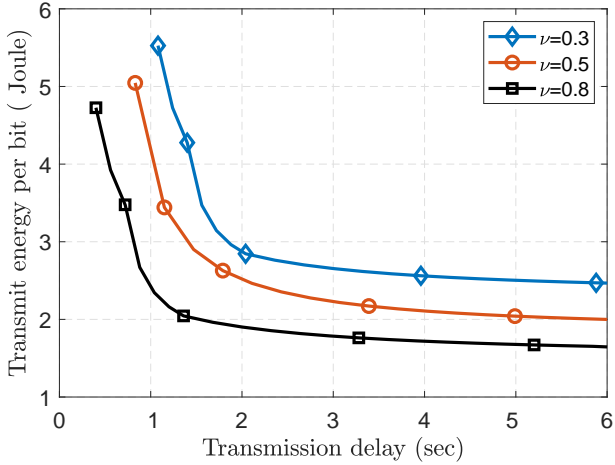}\caption{Energy and \textcolor{black}{Delay trade-off for }the $\mathit{PA}$
scheme \label{fig:- 2}}
\end{figure}

Fig. \ref{fig:- 2} illustrates optimal trade-off curves between energy
and delay corresponding to various values of $\overset{*}{\alpha_{i}}(w_{\mathit{j}})$.
Each point on the curve corresponds to different energy delay levels,
and it is calculated by adjusting $\overset{*}{\alpha_{i}}(w_{\mathit{j}})$
within the range between 0 and 1. Based on the trade-off relationship
between energy and delay, we find that the energy decreases monotonically
with delay. By increasing $\overset{*}{\alpha_{i}}(w_{\mathit{j}})$,
the optimum delay decreases under the same energy, as the $\overset{*}{\alpha_{i}}(w_{\mathit{j}})$
for the delay objective is decreased. In the same manner, increasing
$\overset{*}{\alpha_{i}}(w_{\mathit{j}})$ leads to reduce energy
consumption due to the long delay in data delivery.\textbf{ }As observed,
any curve associated with a specific $\overset{*}{\alpha_{i}}(w_{\mathit{j}}),$\textbf{
}follows an exponential decay unit approaches to a steady when the
transmission delay leans to infinity. In this case, the energy converges
to a constant as delay approaches $\overset{*}{\alpha_{i}}(w_{\mathit{j}})$
constraint value. The energy, on the other hand, approaches $\overset{*}{\alpha_{i}}(w_{\mathit{j}})$
constraint value when the delay tends to infinity. This result is
consistent with earlier researchers results such as \cite{Gurakan2016},
which reveals that there is always a trade-off between energy and
delay that enables the performance of the transmission network to
be enhanced. As a continuation of these studies, our proposal in $\mathit{PA}$
provides a novel trade-off scheme, adopting an energy allocation strategy
to achieve the optimal energy distribution between relay and user
nodes, which, in turn, improves UAV transmission network performance
by reducing the transmission delay (i.e., high data rate) or energy
consumption (i.e., higher energy efficiency). 

\begin{figure}
\centering{}\includegraphics[width=3.5in]{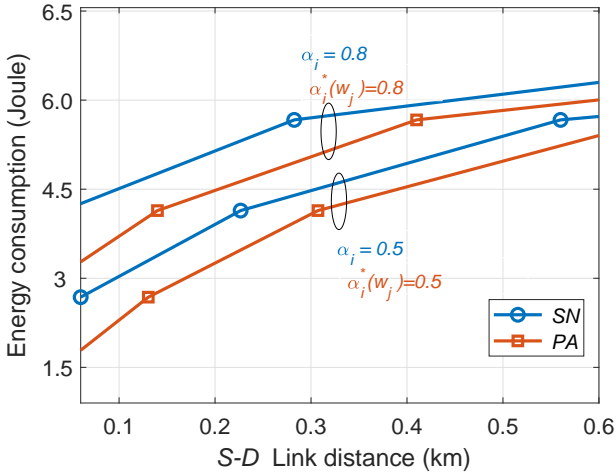}\caption{Energy consumption \textcolor{black}{vs link distance for }the $\mathit{PA}$
and $\mathit{SN}$ schemes \label{fig:- 4}}
\end{figure}

\begin{figure}
\centering{}\includegraphics[width=3.5in]{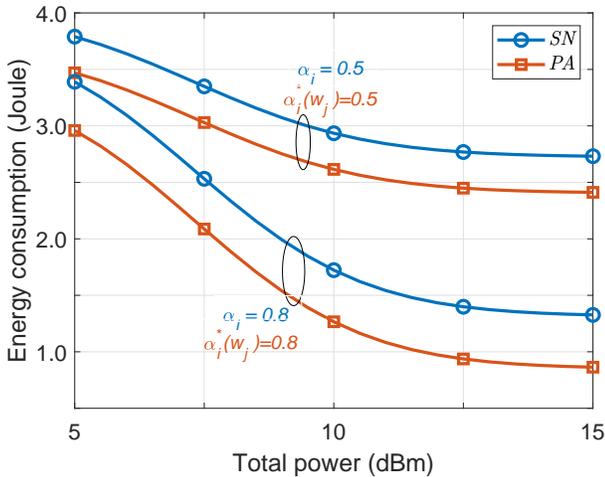}\caption{Energy consumption \textcolor{black}{vs }total power \textcolor{black}{for
}the $\mathit{PA}$ and $\mathit{SN}$ schemes\label{fig:- 5}}
\end{figure}

Consider the following practical example to illustrate how the framework
proposed can be implemented for UAV networks: transmitting signals,
by Sa or Sb, over poor channel conditions consumes more energy than
broadcasting over normal channel conditions. Under such a scenario,
using the proposed method enables a user's node to determine whether
to send packets or leave the users idle based on the energy consumption.
Then, the proposed approach ensures that energy consumption is conserved
in UAV networks by creating a decision-making scheme. In order to
clarify the characteristic of this decision-making platform, Fig.
\ref{fig:- 4} illustrates the energy consumption  curves versus total
distance $d$, which represents the sum of $d_{a}$ and $d_{b}$.
It is clearly shown that $\mathit{d}$ has a significant impact on
energy consumption output as it decreases significantly with increasing
$\mathit{d}$. This is because increasing $d_{a}$ requires a higher
transmit power by $\mathit{S_{a}}$ node to suppress channel-fading
growth, and, hence, higher energy consumption. This result agrees
with what is observed in\textbf{ }\cite{Yang2016}. In addition, Fig.
\ref{fig:- 4} shows that the proposed scheme allows energy consumption
to be decreased by about 15\% more than the $\mathit{SA}$ scheme. 

\begin{figure}
\centering{}\includegraphics[width=3.5in]{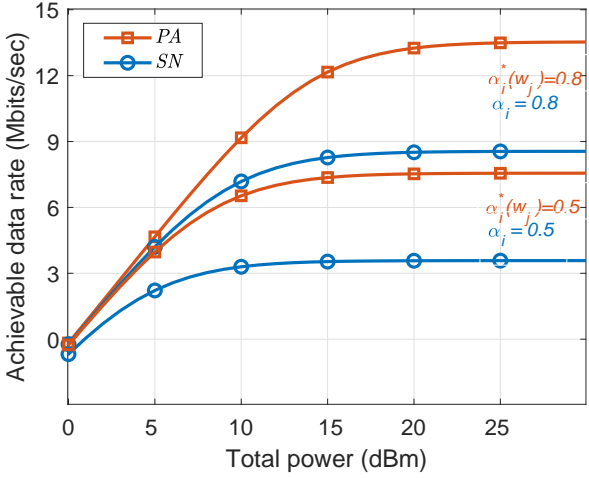}\caption{Achievable data rate \textcolor{black}{vs }total power\textcolor{black}{{}
for }the $\mathit{PA}$ and $\mathit{SN}$ schemes\label{fig:- 6}}
\end{figure}

\begin{figure}
\centering{}\includegraphics[width=3.5in]{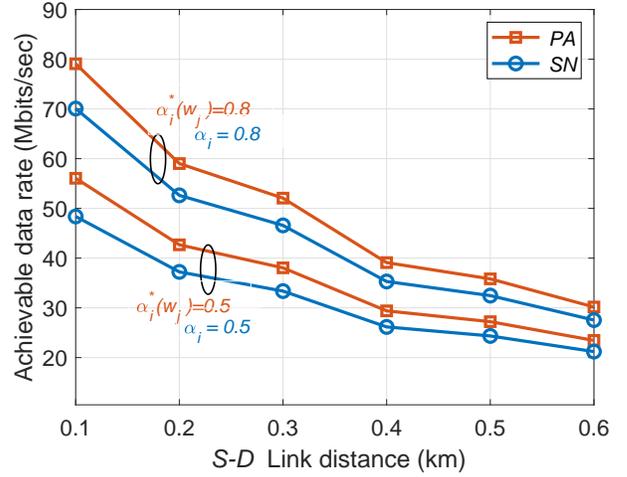}\caption{Achievable data rate \textcolor{black}{vs link distance for }the $\mathit{PA}$
and $\mathit{SN}$ schemes \label{fig:- 7}}
\end{figure}

Fig. \ref{fig:- 5} depicts that with the rise in power, $\mathcal{E}$
initially decreases and then reaches a constant. This is because $\mathcal{R}$
only increases logarithmically with $\mathrm{\mathit{\overline{P}}}$
while the transmission power consumption increases linearly with the
transmission power \cite{Xiong2012}. Consequently, increasing power
divides energy consumption into two stages. In the first stage, the
energy is decreased; then, in the second stages, it approaches a constant
rate. This reveals that there is a \textit{transition point} that
allows for the improvement of energy consumption.\textbf{ }Fig. \ref{fig:- 5},
illustrates that the domain of the transition point is between 13
and 17 dBm; and a further reduction of energy consumption is shown
by the proposed method of $\mathit{PA,}$ which corresponds to previous
studies \cite{Xiong2012,Ng2012}. Further validation is given by Fig.
\ref{fig:- 6},\textbf{ }which compares the average data rate and
the maximum transmit power. It is clear that the average data rate
in \eqref{eq: opt_qt} tends to become a constant at the high range
of the transmit power (range between 14 and 16 dBm). This is because
the power allocation strategy enables the transmit power to be regulated
\cite{Ng2012a}. By using the $\mathit{PA}$ scheme,\textbf{ }a higher
average system rate is achieved, as the $\mathit{PA}$ minimises both
$\mathcal{E}$ and $\mathit{q}$ (i.e., increases $\mathcal{R}$)
simultaneously. Therefore, $\mathcal{R}$ increases logarithmically,
while the transmission power consumption increases linearly with the
transmission power \cite{Xiong2012}. In other words, the transition
point is improved significantly, and such enhancement is expected
to enhance other performance parameters in UAV networks. For example,
Fig. \ref{fig:- 7} illustrates the relationship between distance
and achievable data rate. It shows that increasing distance between
$\mathit{S_{a}}$ and $\mathit{\mathit{S_{b}}}$ nodes gradually reduces
achievable data rates due to high channel gain degradation.\textbf{
}However, the $\mathit{PA}$ is always better than the $\mathit{SN}$,
which agrees with other results obtained by \cite{Tian2019}. 

\begin{figure}
\centering{}\includegraphics[width=3.5in]{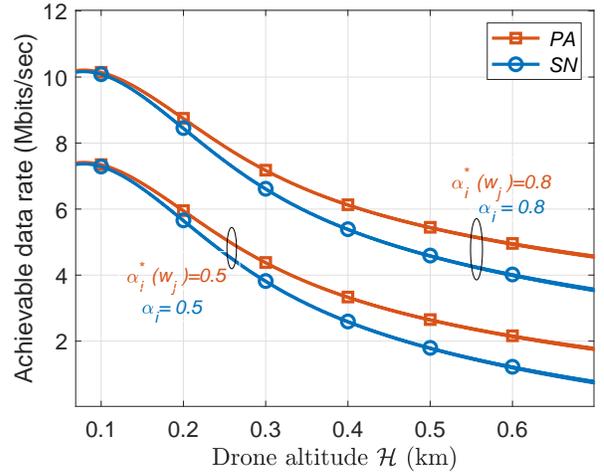}\caption{Achievable data rate \textcolor{black}{vs }altitude flight\textcolor{black}{{}
for }the $\mathit{PA}$ and $\mathit{SN}$ schemes \label{fig:- 8}}
\end{figure}

\begin{figure}
\centering{}\includegraphics[width=3.5in]{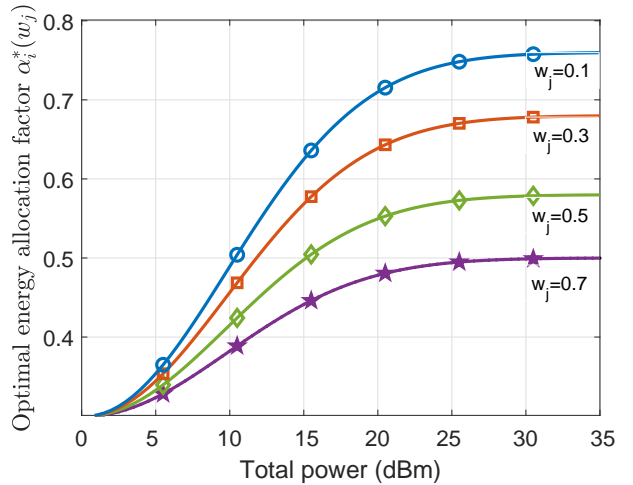}\caption{Achievable data rate \textcolor{black}{vs }total power\textcolor{black}{{}
for }the $\mathit{PA}$ scheme\label{fig:-9}}
\end{figure}

In the $\mathit{PA}$, the rate of distance change depends on flight
altitude, and this is required to explain the effect of the drone
altitude on data rate performance, as illustrated in Fig. \ref{fig:- 8}.
As depicted in the figure, when the drone takes off from the $\mathit{S_{a}}$
node level at 100 m, the data rate decreases slightly with the distance
travelled by the drone remains within a low path loss range \cite{Khawaja2019}.
By increasing the altitude between $\mathit{S_{a}}$ node and drone,
the path loss increases significantly and the data rate, in turn,
starts to decrease gradually. The results of the proposed algorithm
in the $\mathit{PA}$ show a marked improvement because increasing
the altitude extend $d_{a}$ and, then, the arrival data rate at the
relay is decreased. Hence, in addition to reducing transmission delay,
both user nodes allocation of a higher power to each other\textbf{
}in order to motivate information growth. The data rate curve is enhanced
compared with other studies that depend on energy allocation only,
as in \cite{Sboui2017}. To explain how the proposed $\overset{*}{\alpha_{i}}(w_{j})$
is allocated for the UAV network, Fig. \ref{fig:-9} plots the relationship
between $\overset{*}{\alpha_{i}}(w_{j})$ and both $\overline{P}$
and $w_{j}.$ It can be seen that increasing $\overline{P}$ allows
$\overset{*}{\alpha_{i}}(w_{j})$ to be increased while notably decreasing
$w_{j}$. The user nodes manage such a high $\overline{P}$ based
on many factors, such as channel conditions or distance $d_{a}$ and
$d_{b}$. Thus, if $d_{a}$ is higher than $d_{b},$ the path loss
of $d_{a}$ is large and the rate at which data arrives at the relay
is low. Therefore, the user nodes allocate higher $\overset{*}{\alpha_{i}}(w_{j})$
to stimulate an increase in data transmission by each user when there
is an increase in delay time.

\begin{figure}
\centering{}\includegraphics[width=3.5in]{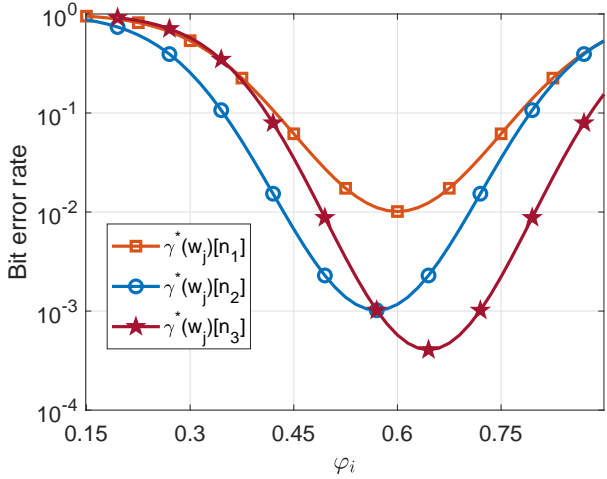}\caption{Power allocation \textcolor{black}{vs }bit error rate\textcolor{black}{{}
for }the $\mathit{PA}$ scheme\label{fig:- 10}}
\end{figure}

To evaluate UAV performance in \eqref{eq: BER}, Fig. \ref{fig:- 10}.
shows results of $\text{\texthtb}_{e}[n]$ versus $\varphi_{i}$,
and for various $\mathit{\textit{\textit{SNRs}}}$ values, where $\overset{*}{\gamma}(w_{j})[n_{3}]>\overset{*}{\gamma}(w_{j})[n_{2}]>\overset{*}{\gamma}(w_{j})[n_{1}].$
Each $\overset{*}{\gamma}(w_{j})[n]$ is obtained from a specific
value of $\varphi_{i}$ ($0<\varphi_{i}<1).$ It is clearly seen that
the highest $\mathit{\textit{\textit{SNR}}}$, at $\overset{*}{\gamma}(w_{j})[n_{3}]$,
enhanced the bit error rate result, and this result confirms the theoretical
analysis, as high $\mathit{\textit{\textit{SNR}}}$ calculates from
optimal power allocation. This result agrees with previous studies
\cite{Mondelli2012}, which indicate that the power allocation $\varphi_{i}$
is an effective metric for enhancing UAV performance.

\begin{figure}
\centering{}\includegraphics[width=3.5in]{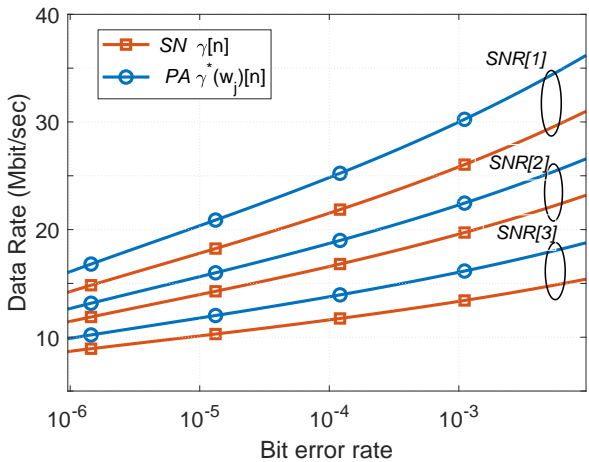}\caption{Achievable data rate \textcolor{black}{vs }bit error rate\textcolor{black}{{}
for }the $\mathit{PA}$ and $\mathit{SN}$ schemes\label{fig:-11}}
\end{figure}

Fig. \ref{fig:- 10}. compares between the $SN$ and $PA$ system
in term of $\varphi_{i}$ and bit error rate. Using energy allocation
in \eqref{eq: Opt-Ei} rises $\mathit{\textit{\textit{SNR}}}${[}$\mathit{n}${]}
which in turn increases data rate, and the result enhancing bit error
rate. Fig. \ref{fig:-11}. also shows that the highest $\mathit{\textit{\textit{SNR}}}$,
i.e. $\textit{\textit{SNR}}[n_{3}]>\textit{\textit{SNR}}[n_{2}]>\textit{\textit{SNR}}[n_{1}]$
$\textit{\textit{SNR}}[n]\in\left\{ \gamma(w_{j})[n],\overset{*}{\gamma}(w_{j})[n]\right\} ,$
gives the lower bit error rate. This result is consistent with previous
results \cite{AlHanafy2012} for terrestrial communication systems.
Obviously, a higher $\mathit{\textit{\textit{SNR}}}$ results in better
performance for the $PA$ as compare in $\mathit{SN}$, consequently,
the proposed $\overset{*}{\text{\texthtb}}[n]$ can improve UAV performance.\textbf{ }

\section{Conclusion and Future Work\label{sec:Conclusion}}

This paper offers a new method for drawing out the most effective
energy-delay curve for a UAV layout by optimising the energy allotments
for users and relay nodes. A multi-objective technique for optimising
the energy allocation factors, weight scalarization optimisation,
is appraised in this paper. The signal transmission distance is taken
into consideration to evaluate UAVs. Simulations confirm the results
originating from analytical expressions, and a real-world application
scenario demonstrates how the proposed structure will be used. The
paper concludes that the proposed technique effectively executes optimal
decision-making and presents a compromise between energy and delay
in UAVs. It would be interesting to extend our proposed study by considering
future works' weighted interval scheduling problem. Also, different
system scenarios such as multiple UAVs could be employed instead of
a single UAV. Besides that, statistical channel state information
estimation can be used instead of instantaneous channel state information.
Further, computational complexity is another direction that would
be required to evaluate in future studies. 

{\footnotesize  

\bibliographystyle{IEEEtran}
\addcontentsline{toc}{section}{\refname}\bibliography{MainRef}

}
\end{document}